\documentclass[%
 aip,
 amsmath,amssymb,
 preprint,%
]{revtex4-1}

\usepackage{graphicx}
\usepackage{dcolumn}
\usepackage{bm}

\usepackage[utf8]{inputenc}
\usepackage[T1]{fontenc}
\usepackage{mathptmx}
\usepackage{etoolbox}
\usepackage{mathtools}
\usepackage{tikz}
\usetikzlibrary{shapes.geometric, arrows}
\usetikzlibrary{arrows.meta}

\makeatletter
\def\@email#1#2{%
 \endgroup
 \patchcmd{\titleblock@produce}
  {\frontmatter@RRAPformat}
  {\frontmatter@RRAPformat{\produce@RRAP{*#1\href{mailto:#2}{#2}}}\frontmatter@RRAPformat}
  {}{}
}%
\makeatother

\usepackage{float}

\begin{document}
\title{Stochastic Multi Configuration Time-Dependent Hartree for Dissipative Quantum Dynamics with Strong Intramolecular Coupling}
\author{Souvik Mandal$^\ast$}
 \email[corresponding author:]{souvikmandal1989@gmail.com}
\affiliation{ 
  Laboratoire de Physique et Chimie Th{\'e}oriques,
  UMR 7019 CNRS/Universit{\'e} de Lorraine,
  1 Bd Arago, 57070 Metz, France
}%

\author{Fabien Gatti}
\affiliation{%
  Institut de Sciences Mol{\'e}culaires d'Orsay,
  UMR 8214 CNRS/Universit{\'e} Paris-Saclay,
  B{\^a}t 520, rue Andr{\'e} Rivi{\`e}re, 91405 Orsay CEDEX, France
}%

\author{Oussama Bindech}
\author{Roberto Marquardt}%
\affiliation{Laboratoire de Chimie Quantique - Institut de Chimie -
{UMR~7177~CNRS/Unistra}\\
Universit\'e de Strasbourg\\
4, rue Blaise Pascal - CS 90032 - 
67081 STRASBOURG CEDEX - France}

\author{Jean Christophe Tremblay$^\ast$}%
 \email[corresponding author:]{jean-christophe.tremblay@univ-lorraine.fr}
\affiliation{ 
  Laboratoire de Physique et Chimie Th{\'e}oriques,
  UMR 7019 CNRS/Universit{\'e} de Lorraine,
  1 Bd Arago, 57070 Metz, France
}%

\date{\today}

\begin{abstract}
In this article, we explore the dissipation dynamics of a strongly coupled multidimensional system in contact with a Markovian bath following a system-bath approach. We use in this endeavour the recently developed stochastic Multi-Configuration Time-Dependent Hartree approach within the Monte Carlo wave packet formalism [J.\,Chem.\,Phys.\,156, 094109 (2022)]. The method proved to yield thermalized ensembles of wave packets when intramolecular coupling is weak.
To treat strongly coupled systems, new Lindblad dissipative operators are
constructed as linear combinations of the system coordinates and associated momenta. These are obtained by an unitary transformation to a normal mode representation, which reduces intermode coupling up to second order. Additionally, we use combinations of generalized raising/lowering operators to enforce the Boltzmann distribution in the dissipation operators, which yield perfect thermalization in the harmonic limit.
The two ansatz are tested using a model two-dimensional hamiltonian parameterized to disentangle the effects of intramolecular potential coupling, of strong mode mixing observed in Fermi resonances, and of anharmonicity.
\end{abstract}

\maketitle

\section{Introduction}

The study of the quantum dynamics of multidimensional systems is an extremely demanding task, both experimentally and theoretically. From an experimental point of view, the isolation of features that are characteristic of the quantum mechanics of the underlying processes is a crucial aspect of measurements that often suffer from perturbing
 incoherent processes. The main theoretical difficulty originates from the nature of the wavefunctions which are typically spread over a large number of degrees of freedom and therefore require a large numerical basis to be characterized. If the number of strongly interacting particles is very large, the standard solutions of the quantum mechanical equations of motion become prohibitively expensive with presently available computational facilities because of the highly increased density of states. However, often we are interested in understanding the dynamics of a certain subsystem of strongly interacting particles, which interact weakly with a large environment. This situation then becomes more manageable within a so-called system-bath~\cite{breuerpetrobook,nitzan2006chemical,blum2012density} approach, where the system is tackled full quantum mechanically, while the bath acts as a reservoir for energy transfer. 

In a previous work,~\cite{Mandal2022sMCTDH} we examined the possibility of describing an open quantum system with energy dissipation and population relaxation in connection to a thermostat within the Lindblad formalism. This new method complements other work in high dimensional system-bath dynamics.~\cite{Gao:1997, Gao:1998,nes03:24,nest2000open,nest2002improving,pesce1997free,pesce1998coupled,pesce1998variational,chemrev,tremblay2008selective,tremblay2009selective,10:TMS:diss,11:TMS:noau,tremblay2011laser,tremblay2012excitation,fuchsel2012selective,tremblay2013unifying,Tremblay:2019b} In our approach, a Markovian master equation~\cite{Gao:1998,Ford:1999} is derived for a Lindblad operator in sum-of-product form. The master equation is solved by propagation of stochastic wave packets~\cite{Breuer:2007} using the Monte Carlo Wave Packet (MCWP) methodology,\cite{dalibard1992wave,dum1992monte,carmichael1993,molmer1993monte} which is now implemented in the Heidelberg Multi-Configuration Time-Dependent Hartree (MCTDH) code.\cite{mctdh:package} The method was benchmarked for a two-dimensional vibrational
model system. The population of different states was found to behave thermally within their statistical fluctuations in the weak intramolecular coupling regime. For strong coupling, a hyperthermal behaviour was observed, which originates from both intramolecular coupling and anharmonicity, and which lacks a sound physical interpretation.

In this paper, we present solutions to overcome the unphysical hyperthermal behaviour. First, we introduce a linear coordinate transformation to reduce the potential coupling upto second order, similarly to a normal mode picture. Secondly, we introduce generalized raising/lowering operators to enforce thermal behaviour more pertinently. The paper is organized as follows. In Section \ref{theory}, the dissipative operators and their rotation to normal coordinates are discussed. Section \ref{model}  describes the different model systems used to benchmark the new approach. In Section \ref{result}, the thermalization dynamics of the different model systems is studied. Section \ref{conclusion} presents our conclusions.

\section{Theory}\label{theory}
We have recently implemented the Monte Carlo Wave Packet (MCWP) method~\cite{dalibard1992wave,dum1992monte,carmichael1993,molmer1993monte} in the Heidelberg Multi-Configuration Time-Dependent Hartree (MCTDH) code~\cite{mctdh:package} to study dissipative quantum dynamics. It is a stochastic unraveling of the reduced density matrix propagation, which we dubbed sMCTDH for ``stochastic MCTDH''. In a nutshell the quantum dynamics of a system in contact with a thermal bath is represented by a set of solutions of a stochastic  time dependent Schr\"{o}dinger equation. These solutions are called \textit{realizations}. Each stochastic wavepacket is propagated piecewise deterministically\cite{breuerpetrobook} under the operation of a nonlinear hamiltonian, which contains a dissipative term. The environment induces fluctuations in the system, which we account for by  sampling the distribution of waiting times. On average over many realizations of the stochastic process, we recover the ensemble dynamics of the reduced density matrix. The details of the MCWP method and its algorithm can be found elsewhere.~\cite{Tremblay:2019b, Mandal2022sMCTDH}

Here we only focus on the specific form of the nonlinear hamiltonian chosen for the sMCTDH method   
\begin{eqnarray}
{\hat H}&=&{\hat H}_{\rm sys} -\frac{{\rm i} \hbar}{2}\sum\limits_{k=1}^f \sum\limits_{j=1}^{M_{\rm op}} \hat{A}_{kj}^{\dagger}(T) \hat{A}_{kj}(T) 
\label{total-ham}
\end{eqnarray}
where $\hat{H}_{\rm sys}$ is the system hamiltonian, the specific form of which  will be discussed in Section \ref{model}. $\sum\limits_{j=1}^{M_{\rm op}} \hat{A}_{kj}^{\dagger}(T)
\hat{A}_{kj}(T)$ is a temperature-dependent dissipation operator with bath temperature $T$, and $\hbar$ is the reduced Planck constant. $f$ is the total number of degrees of freedom and $M_{\rm op}$ is the number of operators for each mode $k$. This factorizable form of the dissipation operator is also consistent with the form of the MCTDH wavefunction. We chose $M_{\rm op}$ = 3 for each mode $k$ and the corresponding three different dissipation channels can be defined for each individual mode of the system of interest by the following operators~\cite{Gao:1997, Mandal2022sMCTDH,Ford:1999}
  \begin{equation}
    \hat{A}_{k\,1}(T) = \mu_k(T)\,\hat{w}_k,\qquad 
    \hat{A}_{k\,2}(T) = \nu_k(T)\,\hat{p}_k,\qquad 
    \hat{A}_{k\,3} = \kappa_k\,\left(\hat{w}_k + \frac{\rm i }{m_k\,\omega_k}\,\hat{p}_k\right)\,.
    \label{ZR-relaxOP}
  \end{equation}
Here, $\hat{w}_k$ is the multiplicative operator of a generalized coordinate for mode $k$
and $\hat{p}_k$ is the associated momentum operator. As $\hat{w}$ is multiplicative operator, we may leave out the circumflex over the symbol, when the symbol represents the coordinate. $\omega_k$ is the harmonic angular frequency of the potential considered and $m_k$ is the corresponding reduced mass. The first two operators $\hat{A}_{k1}(T)$ and $\hat{A}_{k2}(T)$ induce fluctuations in the system. The operator $\hat{A}_{k3}$ is the generalized lowering operator for mode $k$. With the above definition, Eq. \eqref{ZR-relaxOP}, the dynamics of a two-dimensional model system consisting of two coupled vibrational modes was investigated in our previous work.~\cite{Mandal2022sMCTDH} In the strong intermode coupling regime a hyperthermal behaviour of the high-energy mode was observed at asymptotic times. 
This hyperthermal behaviour was particularly strong for the case where the high-energy mode relaxes more slowly than the low-energy mode. In our the previous work~\cite{Mandal2022sMCTDH}, we gave convincing evidence that the hyperthermal behaviour arises from intermode coupling. In order to reduce the effect of intermode coupling, we advocate here a linear transformation of the generalized set of coordinates $w_k$ to a \textit{normal mode} representation denoted ${Q}_k$ defined by the following reciprocal relations:
\begin{eqnarray}
Q_k = \sum_{l=1}^f c_{kl} \sqrt{m_l} w_l
\label{normal-mode}
\end{eqnarray}
\begin{eqnarray}
w_l = \sum_{k=1}^f \frac{c_{lk}}{\sqrt{m_l}} Q_k
\label{general-mode}
\end{eqnarray}
For the sake of clarity, we call in this paper the coordinates $w_l$ \textit{local mode coordinates}. The coefficients $c_{kl}$ compose the orthonormal eigenvectors of the hessian matrix 
\begin{eqnarray}
H_{f} = \begin{pmatrix}
\frac{1}{m_1}\frac{\partial^2 V}{\partial w_1^2 } & \frac{1}{\sqrt{m_1 m_2}}\frac{\partial^2 V}{\partial w_1 \partial w_2} & \ldots & \frac{1}{\sqrt{m_1 m_f}}\frac{\partial^2 V}{\partial w_1 \partial w_f} \\
\frac{1}{\sqrt{m_2 m_1}}\frac{\partial^2 V}{\partial w_2 \partial w_1} & \frac{1}{m_2}\frac{\partial^2 V}{\partial w_2^2} & \ldots & \frac{1}{\sqrt{m_2 m_f}}\frac{\partial^2 V}{\partial w_2 \partial w_f}   \\
\vdots & \vdots & \ddots & \vdots \\
\frac{1}{\sqrt{m_f m_1}}\frac{\partial^2 V}{\partial w_f \partial w_1} & \frac{1}{\sqrt{m_f m_2}}\frac{\partial^2 V}{\partial w_f \partial w_2} & \ldots & \frac{1}{m_f}\frac{\partial^2 V}{\partial w_f^2}  
\end{pmatrix}  
\label{hessain}
\end{eqnarray}
where $V$ is the potential energy hypersurface. That allows us to decouple the modes exactly upto second order in the potential expansion. All higher-order couplings will still remain after the transformation. The angular frequency $\omega_k^{\prime}$ for the corresponding normal mode is calculated from the eigenvalue ($\lambda_k$) of $H_f$.
\begin{eqnarray}
\omega^{\prime}_{k} = \sqrt{\lambda_k}
\label{norm-freq}
\end{eqnarray}
 The conjugate momentum of $\hat{Q}_k$ can be defined as
  \begin{equation}\label{norm-momenta}
\hat{p}_{Q_k}  = -{\rm i} \hbar \sum_{l=1}^f \frac{\partial w_l}{\partial Q_k} \frac{\partial}{\partial w_l}
  \end{equation}
The quantity $\frac{\partial w_l}{\partial Q_k} = \frac{c_{lk}}{\sqrt{m_l}}$ is just a number, which can be straightforwardly calculated from the linear transformation, Eq. \eqref{general-mode}. Using the normal mode coordinate $Q_k$ with frequency $\omega^{\prime}_{k}$ and the corresponding conjugate momentum $\hat{p}_{Q_k}$ as defined in Eq. \eqref{normal-mode}, \eqref{norm-freq} and \eqref{norm-momenta}, the dissipation operators can be defined for individual normal modes following Eq. \eqref{ZR-relaxOP}
  \begin{equation}\label{norm-relaxOP}
    \hat{A}_{k\,1}(T) = \mu_k(T)\,Q_k,\qquad 
    \hat{A}_{k\,2}(T) = \nu_k(T)\,\hat{p}_{Q_k},\qquad 
    \hat{A}_{k\,3} = \kappa_k\,\left(Q_k + \frac{\rm i}{\omega^{\prime}_{k}}\,\hat{p}_{Q_k}\right)
  \end{equation}
Since the normal modes are linear combination of mass-weighted 
generalized coordinates, the mass factor is included in the mass-weighted momentum  $\hat{p}_{Q_k}$. Substituting Eq.  \eqref{normal-mode} and Eq. \eqref{norm-momenta} in Eq. \eqref{norm-relaxOP} yields a redefinition of the dissipation operators
 in terms of local mode coordinates
   \begin{eqnarray}\label{relaxOP}
 &&   \hat{A}_{k\,1}(T) = \mu_k(T)\,\sum_{l=1}^f c_{kl} \sqrt{m_l} w_l,\qquad 
    \hat{A}_{k\,2}(T) =  \,-{\rm i} \hbar \nu_k(T)\, \sum_{l=1}^f  \,\frac{c_{lk}}{\sqrt{m_l}} \frac{\partial}{\partial w_l},\nonumber \\
 &&   \hat{A}_{k\,3} = \kappa_k\,\left(\sum_{l=1}^f \left(c_{kl} \sqrt{m_l} w_l + \frac{ {\hbar}}{\omega^{\prime}_{k}}\,  \frac{c_{lk}}{\sqrt{m_l}}\frac{\partial}{\partial w_l} \right)\right)
  \end{eqnarray} 
  We recognize that the dissipation operators in the normal mode representation are simple linear combinations of the same operators in local mode coordinates.
  As such, they retain a simple sum-of-product form compatible with the sMCTDH algorithm. For each mode 3 different mechanisms for energy dissipation can be identified. Altogether, there will be $3\times f$ \textit{dissipation channels} that will give different significance to the overall, effectively temperature-dependent dissipation rate. The coefficients $\mu_k(T)$, $\nu_k(T)$, and $\kappa_k$ are chosen to enforce thermal behaviour in an uncoupled one-dimensional harmonic mode~\cite{Gao:1998,Ford:1999}, and they take the form
\begin{eqnarray}\label{coeffs1}
\mu_k^2(T)  &=&\frac{\gamma_k  \omega^{\prime}_{k}}{2 \hbar}\left(\coth\Big(\frac{\hbar\omega^{\prime}_{k}}{2k_{\rm B}\,T}\Big)-1\right) \\ \label{coeffs2}
\nu_k^2(T) &=&\frac{\gamma_k}{2 \hbar  \omega^{\prime}_{k}}\left(\coth\Big(\frac{\hbar\omega^{\prime}_{k}}{2k_{\rm B}\,T}\Big) -1\right)\\\label{coeffs3}
\kappa_k^2 &=& \frac{\gamma_k \omega^{\prime}_{k}}{2\hbar}
\end{eqnarray}
where $k_{\rm B}$ is the Boltzmann constant and the quantity $\gamma_k$ is a rate parameter. The frequency $\omega_k^{\prime}$ is obtained from Eq. \eqref{norm-freq} and thus it contains the information on the intermode coupling upto second order. For a derivation of Eqs. \eqref{coeffs1}-\eqref{coeffs3}, the reader is referred to the literature~\cite{Mandal2022sMCTDH}. 

In order to achieve thermal distribution exactly in the limit
of the bilinearly coupled harmonic oscillators, the temperature-dependent coefficients $\mu_k(T)$ and $\nu_k(T)$ can be modified as
\begin{eqnarray}\label{BM-coff1}
     \tilde{\mu}k(T)=\kappa_ke^{\hbar \omega_k^{\prime}/2k_{\rm B}T} \\
          \tilde{\nu}_k(T)=-\frac{\rm i}{\omega_k^{\prime}}\kappa_ke^{\hbar \omega_k^{\prime}/2{k_{\rm B}T}} \label{BM-coff2}
     \end{eqnarray}     
With this new definition of the coefficients, the dissipation operators in Eq. \eqref{norm-relaxOP} can be rewritten as
  \begin{equation}\label{thermal-relaxOP}
    \hat{A}_{k\,1}(T) + \hat{A}_{k\,2}(T)= \kappa_k\,e^{\hbar \omega_k^{\prime}/2k_{\rm B}T}\left(\hat{Q}_k - \frac{\rm i}{\omega^{\prime}_{k}}\,\hat{p}_{Q_k}\right) \qquad
        \hat{A}_{k\,3} = \kappa_k\,\left(\hat{Q}_k + \frac{\rm i}{\omega^{\prime}_{k}}\,\hat{p}_{Q_k}\right)
  \end{equation}
It can be recognized that the sum of the first two dissipation operators, $\hat{A}_{k\,1}(T) + \hat{A}_{k\,2}$(T), becomes a generalized raising operator. The rates associated with the upward and downward transitions are now related via the constant $\kappa_k$, and they are constrained to impose thermal detailed balance exactly. Note that the factor of 2 in the exponent is necessary because the rates appear squared in the Lindblad equation. The physical interpretation of the new set of operators as temperature-dependent fluctuations and dissipation terms is blurred, with the expected advantage of reaching thermal equilibrium at asymptotic times in exactly separable cases.

\section{Model}\label{model}

To study the dissipation dynamics and its thermalization behaviour, we chose a two-dimensional (2D) model consisting of two coupled oscillators as in our previous work~\cite{Mandal2022sMCTDH}. The model is parameterized such as to represent a dioxygen molecule adsorbed on a Pt(111) surface~\cite{Gland:1980,Gao:1998}, where the O-O and the Pt-O$_2$ bonds are the two coupled oscillators.  The O-O vibration ($R$-mode) will be considered to be a harmonic mode, and the O$_2$-Pt vibration ($Z$-mode) will be either a harmonic mode or an anharmonic mode. The $R$ and $Z$-modes are represented by the coordinates $y_R$ and $y_Z$, respectively. We investigate three different types of model potentials for this O$_2$-Pt system below.  

\subsection{Anharmonic strong coupling potential}
We call \textit{anharmonic strong coupling potential} a potential in which the $Z$-mode is anharmonic, the $R$-mode is harmonic and the modes are bilinearly  coupled. The hamiltonian of such a system can be written as
\begin{eqnarray}
\hat{H}_{\rm sys}= \left(-\frac{\hbar^2}{2m_R}\frac{\partial^2}{\partial R^2}\right) + \left(-\frac{\hbar^2}{2m_Z}\frac{\partial^2}{\partial Z^2}\right) + D_{\rm e} \bigg(a_Z^2y_Z^2\,+\,a_R^2\left(y_R\;+\;Cy_Z\right)^2\bigg)
 \label{eq:anharmonic_ham}
\end{eqnarray}
where the last term is the potential $V(R, Z)$. $y_Z$ is a Morse coordinate and $y_R$ is a harmonic coordinate: 
\begin{eqnarray}
  y_Z = \frac{1-{\rm e}^{-a_Z\,(Z-Z_{\rm e})}}{a_Z}, \qquad
  y_R = R-R_{\rm e}
\end{eqnarray}
The quantities $a_Z$ and $a_R$ are defined as
\begin{eqnarray}
  a_Z = \sqrt{\frac{m_Z\,\omega_Z^2}{2\,D_{\rm e}}} \;\;\;\;\;\;{\rm and}\;\;\;\;\;\;
  a_R = \sqrt{\frac{m_R\,\omega_R^2}{2\,D_{\rm e}}}  
\end{eqnarray}
where $D_{\rm e}$ is the depth of the potential well. Here, $R$ is the distance between the two oxygen atoms, with $R_{\rm e}$ being its value at equilibrium. $Z$ is the distance from the center of mass of the oxygen molecule to a platinum atom, with $Z_{\rm e}$ being the equilibrium distance. $\omega_{Z/R}$ and $m_{Z/R}$ are the corresponding frequency and mass of the respective modes. The quantity $C$ is a dimensionless coupling constant. As we focus on the thermalization behaviour in the strong coupling regime the value of the constant $C$ is chosen as 0.5.~\cite{Mandal2022sMCTDH} The reader is referred to ref. \citenum{Mandal2022sMCTDH} for results with moderate and weaker couplings. For this anharmonic strong coupling potential, the local mode operators
 used to define the dissipative operators in Eq.\,\eqref{relaxOP} with $k=\{1,2\}$ take the form 
\begin{equation}\label{local-anhar-coord}
w_1 \equiv w_Z = y_Z - \left(\frac{\hbar}{4 m_Z \omega_Z \Lambda_Z}\right)^{1/2};\qquad 
w_2 \equiv w_R = y_R \\ 
 \end{equation}
 The quantity $\Lambda_k$ is defined by $\Lambda_k = 2D_{\rm e}\,/\,(\hbar\,\omega_k)$.
The eigenstates of $\hat{H}_{\rm sys}$ in 
Eq. \eqref{eq:anharmonic_ham} will be denoted by the ket $\lvert n_{v_Z\,v_R}\rangle$, where $n=0,1,2,\ldots$ is a
  cardinal number specifying the energetic order of the states, while $v_Z$ and
  $v_R$ signify the quantum numbers of those zeroth-order  ($C$ = 0 case)
  states which have the strongest participation in the given eigenstate. The set of the 
  first twelve zeroth-order states and eigenstates is given in Table \ref {tab:anharmonic-energy}.  They are computed using the Heidelbebrg MCTDH package. A basis of 5 single particle functions is used for each mode. The integration is performed on a primitive grid composed of 64 discrete variable functions (sine-DVR) for each one of the two modes.

\begin{table}[htb!]
\begin{center}
\begin{tabular}{@{\hspace*{3mm}}c@{\hspace*{3mm}}|@{\hspace*{3mm}}c@{\hspace*{3mm}}|c@{\hspace*{3mm}}}
\hline\hline 
State & \multicolumn{2}{c} {Relative energies in units of $hc$ cm$^{-1}$} \\
\hline
      & {zeroth-order}           & {eigenstates } \\

& $C = 0$  &$C = 0.5$ \\ 
\hline 
$\lvert0_{00}\rangle$ & 000.0  &000.0 \\ 
\hline 
$\lvert1_{10}\rangle$ & 399.5 &384.4 \\ 
\hline 
$\lvert2_{20}\rangle$ &  770.7 & 739.6\\ 
\hline 
$\lvert3_{01}\rangle$ & 870.0  & 910.5\\ 
\hline 
$\lvert4_{30}\rangle$ & 1113.5 &  1071.9\\ 
\hline 
$\lvert5_{11}\rangle$ & 1269.5 &  1289.6\\ 
\hline 
$\lvert6_{40}\rangle$ & 1427.0 &  1385.4\\ 
\hline 
$\lvert7_{21}\rangle$ & 1640.7 & 1617.1\\ 
\hline 
$\lvert8_{50}\rangle$ & 1713.9 &  1690.3\\ 
\hline 
$\lvert9_{02}\rangle$ &  1740.0&  1819.2\\ 
\hline 
$\lvert10^\ast\rangle$ & 1971.5 &  1894.8\\ 
\hline 
$\lvert11^\ast \rangle$ & 1983.5 & 1995.2\\ 
\hline\hline 
\end{tabular} 
\end{center}
\caption{ Ordering and energies of zeroth-order states and eigenstates of the anharmonic strong coupling potential system (see Eq. \eqref{eq:anharmonic_ham}).
All reported energies are relative to the respective ground vibrational state in units of $hc$ cm$^{-1}$. The states 10 and 11 with the $^\ast$ symbol constitute a Fermi resonance pair between the two zeroth-order states $\lvert v_Z=6,v_R=0\rangle$ and $\lvert v_Z=3,v_R=1\rangle$. The parameters used in Eq. \eqref{eq:anharmonic_ham} are $m_Z$ = 27.48 u, $m_R$ = 7.9995 u,    $\hbar \omega_Z$ = 53 meV, $\hbar \omega_R$ = 108 meV, $Z_e$ = 2.11 \AA, $R_e$ = 1.37 \AA  and $D_{\rm e}$ = 0.4 eV. The parameters are taken from refs.~\citenum{Gao:1998}, ~\citenum{Gland:1980}, \citenum{Menzel:1990} and \citenum{Steiniger:1982}.}
\label{tab:anharmonic-energy}
\end{table} 

\subsection{Fermi resonance potential}
\begin{table}[htb!]
\begin{center}
\begin{tabular}{@{\hspace*{3mm}}c@{\hspace*{3mm}}|@{\hspace*{3mm}}c@{\hspace*{3mm}}|@{\hspace*{3mm}}c@{\hspace*{3mm}}|@{\hspace*{3mm}}c@{\hspace*{3mm}}|c@{\hspace*{3mm}}}
\hline\hline 
 \multicolumn{5}{c} {Relative energies in units of $hc$ cm$^{-1}$} \\
\hline
 \multicolumn{1}{c}  {zeroth-order} & & \multicolumn{3}{c} {eigenstates } \\
\hline
State & $C = 0$ & State &$C = 0.1$ &$C = 0.5$ \\ 
\hline 
$\lvert{0,0}\rangle$ & 000.0 & $\lvert 0_1 \rangle$ & 000.0 &000.0 \\ 
\hline 
$\lvert{1,0}\rangle$ & 400.0 & $\lvert  (1/2)_1 \rangle$ & 400.0 &400.0 \\ 
\hline 
$\lvert{2,0}\rangle$ & 800.0 & $\lvert 1_2\rangle$ & 799.2 & 796.0\\ 
\hline 
$\lvert{0,1}\rangle$ & 800.0 & $\lvert 1_1 \rangle$ & 800.8 & 804.0\\ 
\hline 
$\lvert{3,0}\rangle$ & 1200.0 & $\lvert (3/2)_2 \rangle$  & 1198.6 & 1193.1\\ 
\hline 
$\lvert{1,1}\rangle$ & 1200.0 & $\lvert (3/2)_1 \rangle$ & 1201.4 & 1207.0\\ 
\hline 
$\lvert{4,0}\rangle$ & 1600.0 & $\lvert 2_3 \rangle$ & 1597.7 & 1588.8\\ 
\hline 
$\lvert{2,1}\rangle$ & 1600.0 & $\lvert 2_2 \rangle$ & 1600.0 &1600.0\\ 
\hline 
$\lvert{0,2}\rangle$ & 1600.0 & $\lvert 2_1 \rangle$ & 1602.3 & 1611.4\\ 
\hline 
$\lvert{5,0}\rangle$ & 2000.0 & $\lvert (5/2)_3 \rangle$ & 1996.8 & 1984.1\\ 
\hline 
$\lvert{3,1}\rangle$ & 2000.0 & $\lvert (5/2)_2 \rangle$ & 2000.0 & 2000.1\\ 
\hline 
$\lvert{1,2}\rangle$ & 2000.0 & $\lvert (5/2)_1 \rangle$ & 2003.2 & 2016.2\\ 
\hline\hline 
\end{tabular} 
\end{center}
\caption{ Ordering and energies of zeroth-order states and eigenstates of the harmonic Fermi resonance potential system (see Eq. \eqref{eq:fermi_ham}).
All reported energies are relative to the respective ground vibrational state in units of $hc$ cm$^{-1}$. The parameters used in Eq. \eqref{eq:fermi_ham} are  $m_Z$ = 27.48 u, $m_R$ = 7.9995 u, $\hbar \omega_Z$ = 49.59 meV and $\hbar \omega_R$ = 99.18 meV.}
\label{tab:fermi-energy}
\end{table} 
Here we investigate another type of strong coupling, i.e., due to a Fermi resonance. To this end, we use coupled harmonic oscillators both for the $R$- and $Z$-modes with coordinates $y_R = (R-R_{\rm e})$ and $y_Z = (Z-Z_{\rm e})$, respectively. The $Z$-mode is quadratically coupled with the $R$-mode by a dimensionless coupling constant $C$. The hamiltonian of the system can be written as
\begin{eqnarray}
\hat{H}_{\rm sys}= \left(-\frac{\hbar^2}{2m_R}\frac{\partial^2}{\partial R^2}\right) + \left(-\frac{\hbar^2}{2m_Z}\frac{\partial^2}{\partial Z^2}\right) + \left(\frac{1}{2}\,m_Z\omega_Z^2y_Z^2 + \frac{1}{2}m_R\omega_R^2\left( y_R +  \frac{C}{Z_{\rm e}}  y_Z^2     \right)^2 \right)
\label{eq:fermi_ham}
\end{eqnarray}
where the last term is the potential $V(R, Z)$. Here we set $\omega_R$ = 2$\omega_Z$ with the remaining parameters as in the previous example. Since there is a 2:1 ratio between the frequencies of the two modes, due to the coupling, the creation/annihilation of one vibrational quanta in the $R$-mode will indirectly infer the creation/annihilation of two vibrational quanta in the $Z$-mode. To leading order, the coupling is linear in the $R$-mode, but quadratic in the $Z$-mode. Because of the energetic resonance the two modes are strongly coupled even for small value of $C$. Table \ref{tab:fermi-energy} shows the energies of the first twelve zeroth-order states and eigenstates ($C$ = 0.1 and 0.5) of $\hat{H}_{\rm sys}$ defined in Eq. \eqref{eq:fermi_ham}. The zeroth-order states are defined by $\lvert v_Z, v_R\rangle$, where $v_Z$ and $v_R$ are the quantum numbers for the $Z$- and $R$-mode respectively. The eigenstates are labeled using the polyad notation, where the $j^{\rm th}$ state of the $N$ polyad is denoted as $\mid N_j\rangle$. Here, $N=\max({v_R})+\max({v_Z})/2$, where $\max({v_k})$ refers to the maximal quantum number of the
zeroth-order state, $j=1,\ldots,N+1$, if $N$ is integer, or $j=1,\ldots,N+1/2$, if $N$ is half-integer, and $j=1$ designates the eigenstate with the highest energy within the polyad. This notation was used in refs.~\citenum{Quack:1984,Marquardt:1986}.
For the Fermi resonance, the local mode operators in Eq.\,\eqref{relaxOP} take the form (with $k\in \{1,2\}$) 
 \begin{eqnarray}
\hat{w}_1 \equiv \hat{w}_Z = y_Z ;\qquad 
\hat{w}_2 \equiv \hat{w}_R = y_R 
\label{w_harmZR}
 \end{eqnarray}

\subsection{Bilinearly coupled harmonic potential} 
Finally, we explore the bilinear coupled harmonic potential, where both $R$- and $Z$-modes are harmonic coordinates $y_R = (R-R_{\rm e})$ and $y_Z = (Z-Z_{\rm e})$, respectively. The model hamiltonian follows from that of the strongly coupled anharmonic case in Eq. \eqref{eq:anharmonic_ham} in the limit $D_{\rm e}\to\infty$:
\begin{eqnarray}
&&\hat{H}_{\rm sys}= \left(-\frac{\hbar^2}{2m_R}\frac{\partial^2}{\partial R^2}\right) + \left(-\frac{\hbar^2}{2m_Z}\frac{\partial^2}{\partial Z^2}\right) + \left( \frac{1}{2}\,m_Z\omega_Z^2y_Z^2 + \frac{1}{2}m_R\omega_R^2\left( y_R +  C  y_Z     \right)^2 \right)
\label{eq:hamonic_ham}
\end{eqnarray}
where the last term is the potential $V(R, Z)$. The $R$- and $Z$-modes are bilinearly coupled by the dimensionless coupling constant $C$. The bilinearly coupled harmonic system is useful because it can be perfectly decoupled using the normal mode approach. 
\begin{table}[htb!]
\begin{center}
\begin{tabular}{@{\hspace*{3mm}}c@{\hspace*{3mm}}|@{\hspace*{3mm}}c@{\hspace*{3mm}}|c@{\hspace*{3mm}}}
\hline\hline 
State & \multicolumn{2}{c } {Relative energies in units of $hc$ cm$^{-1}$} \\
\hline
      & zeroth-order           &{eigenstates } \\
& $ C = 0$ & $C = 0.5$ \\ 
\hline 
$\lvert 0_{00}\rangle$ & 000.0 & 000.0 \\ 
\hline 
$\lvert 1_{10}\rangle$ & 395.1 & 378.4 \\ 
\hline 
$\lvert 2_{20}\rangle$ &  790.1&  756.8\\ 
\hline 
$\lvert 3_{01}\rangle$ & 869.9 & 908.2\\ 
\hline 
$\lvert 4_{30}\rangle$ & 1185.2 &  1135.3\\ 
\hline 
$\lvert 5_{11}\rangle$ & 1265.1 & 1286.6\\ 
\hline 
$\lvert 6_{40}\rangle$ & 1580.2 & 1513.7\\ 
\hline 
$\lvert 7_{21}\rangle$ & 1660.1 & 1665.1\\ 
\hline 
$\lvert 8_{02}\rangle$ & 1739.9 &  1816.4\\ 
\hline\hline 
\end{tabular} 
\end{center}
\caption{ Ordering and energies of zeroth-order states and eigenstates of the bilinearly coupled harmonic potential system (see Eq. \eqref{eq:hamonic_ham}).
All reported energies are relative to the respective vibrational ground state in units of $hc$ cm$^{-1}$. The parameters used in Eq. \eqref{eq:hamonic_ham} are  $m_Z$ = 27.48 u, $m_R$ = 7.9995 u, $\hbar \omega_Z$ = 53 meV and $\hbar \omega_R$ = 108 meV.}
\label{tab:harmonic-energy}
\end{table} 
Therefore, it reaches well-defined limits that can be used to benchmark the dynamics of the system. The value of $C$ is 0.5, as above, for the investigation in the strong coupling regime. The eigenstates of $\hat{H}_{\rm sys}$ in Eq. \eqref{eq:hamonic_ham} are denoted by $\lvert n_{v_Z\,v_R}\rangle$ as for the case of an anharmonic bilinear strong coupling potential. The energy values of the first nine zeroth-order states and eigenstates are presented in Table \ref{tab:harmonic-energy}. 
In this case, the local mode operators take the same form as in Eq. \eqref{w_harmZR}.

\section{Results and Discussions}\label{result}
We perform sMCTDH simulations with the model potentials described above. The population dynamics and their thermalization behaviours
at different bath temperatures are investigated in the following subsections. 

\subsection{Anharmonic strong coupling potential}\label{anharmonic}

\begin{figure}[htb!]
\centering
\includegraphics[width=0.42\textwidth,angle=-90]{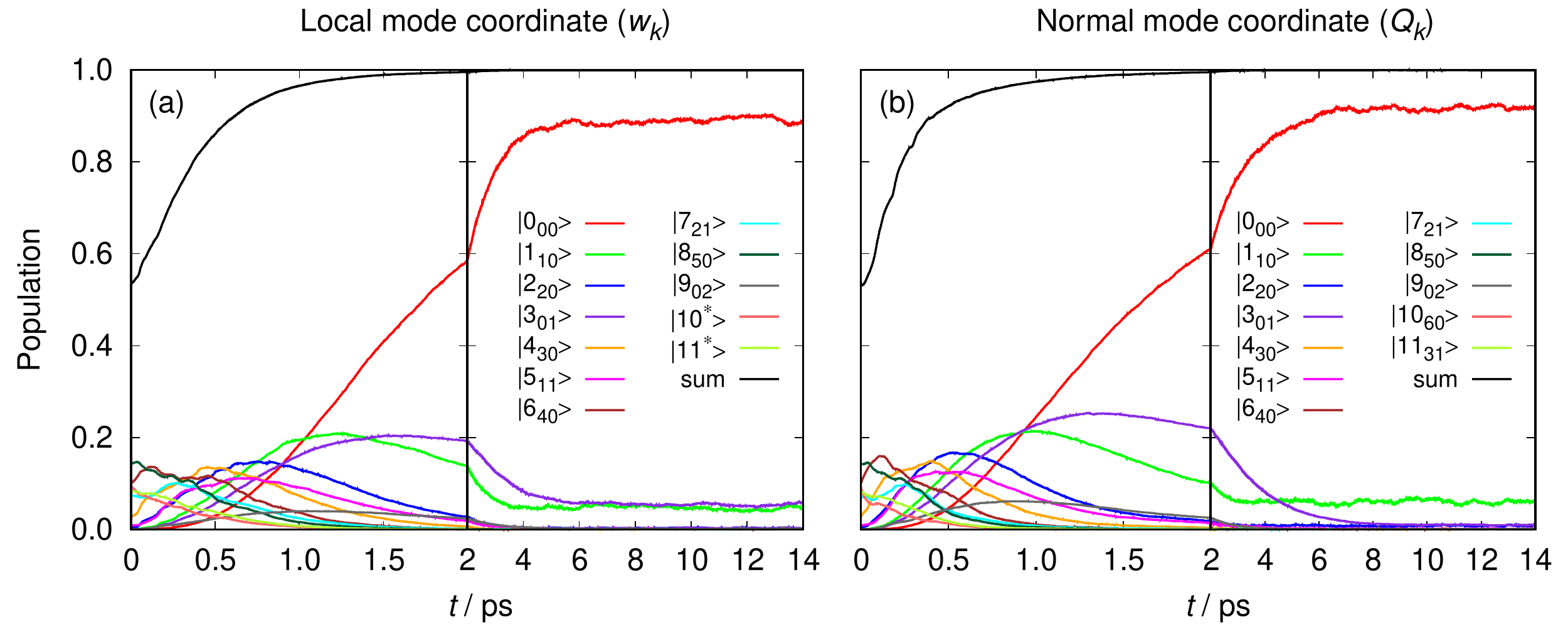}
\caption{Evolution of population of selected eigenstates within the two-dimensional model for O$_2$/Pt(111)  with the anharmonic strong coupling potential (see Eq. \eqref{eq:anharmonic_ham}) at 200 K bath temperature for (a) local mode approach (Eq. \eqref{ZR-relaxOP}) and (b) normal mode  approach (Eq. \eqref{relaxOP}). The initial wavefunction is chosen as a product of a  shifted Gaussian in $Z$ (center = 2.40 \AA, width = 0.039 \AA) 
and a Gaussian in $R$  (center = 1.37 \AA, width = 0.049 \AA).
The relaxation parameter are $\gamma_Z^{-1}$ = 0.5 ps and $\gamma_R^{-1}$ = 2 ps and the coupling strength is $C$ = 0.5. The solid black line indicates the sum of populations of the states which are presented.}
\label{pop_anhar_200K}
\end{figure}

Figure \ref{pop_anhar_200K} presents the population dynamics of the eigensates for the anharmonic strong coupling potential ($C$ = 0.5) described by the system hamiltonian in Eq. \eqref{eq:anharmonic_ham} at a bath temperature of 200 K. The initial wavefunction is chosen as a product of a Gaussian distribution shifted in $Z$ and a Gaussian distribution
 in $R$ centered at $R_{\rm e}$. We  perform the simulation using two different types of dissipation coordinates. Figure \ref{pop_anhar_200K}(a) displays the population dynamics using the dissipation operators as described in Eq. \eqref{ZR-relaxOP}, where the dissipation operators are represented in terms of the local mode coordinates ${w}_k$ defined in Eq. \eqref{local-anhar-coord}.
 On the other hand, Figure \ref{pop_anhar_200K}(b) shows the population dynamics using the dissipation operators as described in Eq. \eqref{relaxOP}.

 As the initial wavefunction is a displaced wave packet along the $Z$ coordinate the initial populations are mostly dominated by the states excited in the $Z$-mode (e.g. states $\lvert 10_{60}\rangle$, $\lvert 8_{50}\rangle$ and $\lvert 6_{40}\rangle$). A few states excited along both the $Z$- and $R$-modes (e.g. states $\lvert 11_{31}\rangle$ and $\lvert 7_{21}\rangle$) are also contributing to the initial population due to strong intermode coupling. With time the population of the excited states transfers to the lower states, mostly via the $Z$-channel because the dissipation rate along the $Z$-mode is faster ($\gamma_Z^{-1}$ = 0.5 ps) compared to that along the $R$-mode ($\gamma_R^{-1}$ = 2 ps ). Finally, the populations go to the ground state, which takes the energy from the bath and re-excites state $\lvert 3_{01}\rangle$ along the $R$-mode.
 The early time dynamics for both the local mode and normal coordinates are mostly similar in nature.  Their behaviour is different at asymptotic times, when the equilibrium is reached. As our interest is on the thermalization of the system, this long-time behaviour will be the focus of the analysis.

 When dissipation is mediated by operators in the local mode representation, Figure \ref{pop_anhar_200K}(a) reveals that the first excited states along each mode reach about the same population.
 This behaviour is in principle inline with a Boltzmann distribution since state $\lvert 3_{01}\rangle$ has a  higher energy than state $\vert 1_{10}\rangle$.
 The overtone of the $Z$-mode, state $\lvert 2_{20}\rangle$, is found at an energy similar to state $\lvert 3_{01}\rangle$, with a separation of about 170 cm$^{-1}$ (see Table \ref{tab:anharmonic-energy}). 
 Consequently, their character become highly mixed by the strong potential coupling.
 This apparently leads to an unphysically high population of the $R$-mode (state  $\lvert 3_{01}\rangle$)
 at asymptotic times, while also suppressing the population of the overtone, state $\lvert 2_{20}\rangle$.
 
This problem is resolved by using the normal mode approach for the dissipation, i.e. when using Eq. \eqref{relaxOP} instead of Eq. \eqref{ZR-relaxOP}, as shown in Figure \ref{pop_anhar_200K}(a). The asymptotic population of state $\lvert 3_{01}\rangle$ which corresponds to equilibrium is found to be strongly reduced. Interestingly, the population of state $\lvert 1_{10}\rangle$ is almost unchanged when moving from the local mode (Eq. \eqref{ZR-relaxOP}) to the normal mode representation, Eq. \eqref{relaxOP}, of the dissipation operators. This could be explained by the fact that state $\lvert 1_{10}\rangle$ ($Z$-mode) exhibits less mixing character even for strong intermode coupling. In Table \ref{tab:anharmonic-energy}, the energetic effect of this strong coupling is nonetheless clearly observed. De facto, at 200 K, the system at equilibrium can be reduced to a two-level system. The temperature associated with this equilibrium can be estimated from the population of the ground state and of state $\lvert 1_{10}\rangle$ at asymptotic times by inverting the Boltzmann distribution and using the energies in Table \ref{tab:anharmonic-energy}.
From Figure \ref{pop_anhar_200K}(a) we find (202 $\pm$ 4) K by averaging the populations over the last two picoseconds, which matches well with the simulation temperature (200 K).

\begin{figure}[htb!]
\centering
\includegraphics[width=0.45\textwidth,angle=-90]{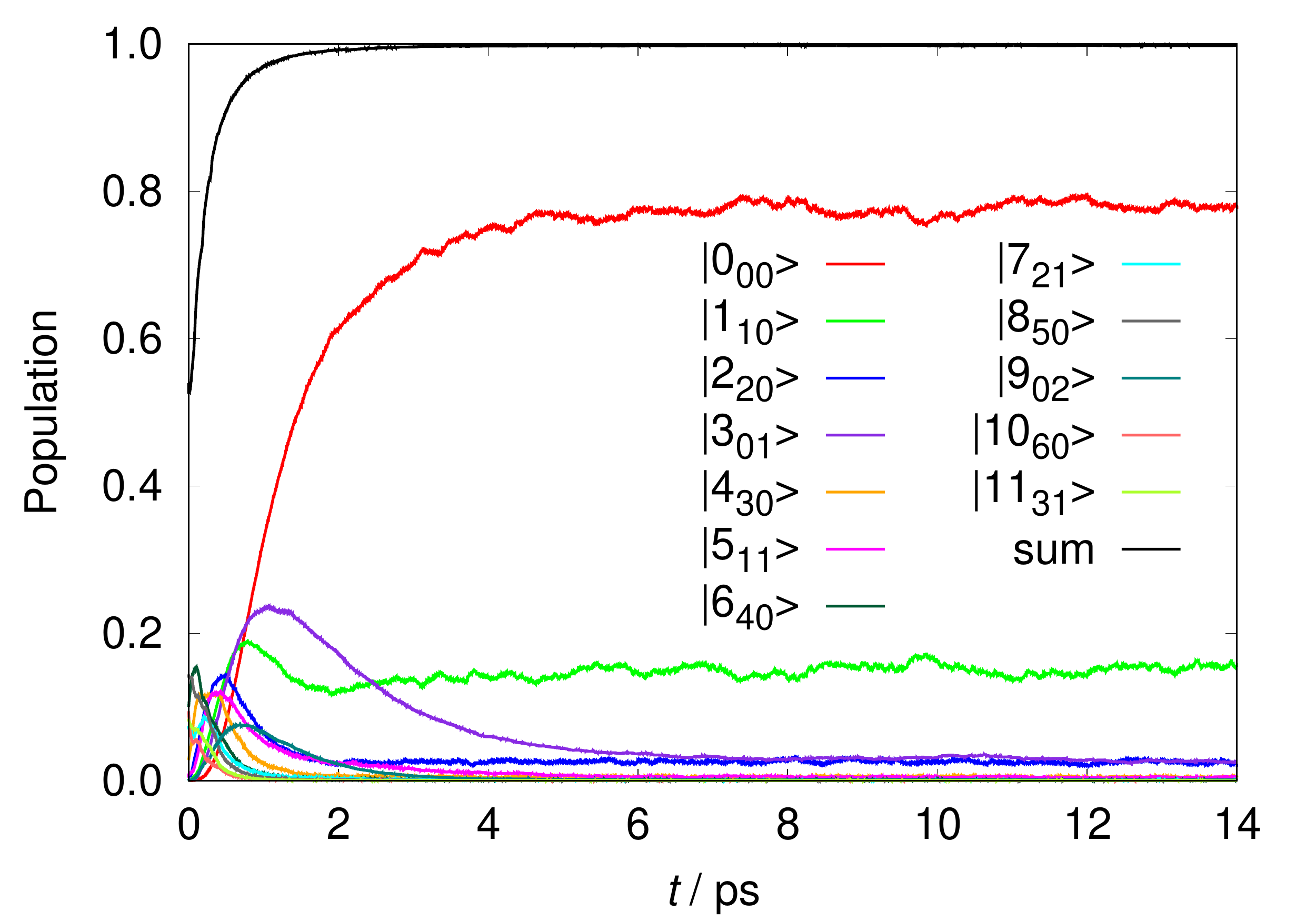}
\caption{Evolution of population of selected eigenstates basis within the two-dimensional model for O$_2$/Pt(111)  with anharmonic strong coupling potential (see Eq. \eqref{eq:anharmonic_ham}) using normal mode approach (Eq. \eqref{relaxOP}) at 400 K bath temperature. The initial conditions, the relaxation parameters and the coupling constant are same as Figure \ref{pop_anhar_200K}.} 
\label{pop_anhar_400K}
\end{figure}

The population dynamics using the normal modes at a bath temperature of 400 K is shown in Figure \ref{pop_anhar_400K} with the same initial condition as in Figure  \ref{pop_anhar_200K}. The populations for different states reach almost constant values at asymptotic times. Here the population of states $\lvert 2_{20}\rangle$ and $\lvert 3_{01}\rangle$ are almost the same. Yet, state $\lvert 2_{20}\rangle$ has lower energy than state $\lvert 3_{01}\rangle$ (see Table \ref{tab:anharmonic-energy}), such that the latter should have a population at equilibrium that is smaller by a factor of about 1.8 from that of the former if the system is to obey thermal detailed balance.
This is an indication that, despite the improvement in the thermalization behaviour resulting from
 the normal mode representation of the dissipation operators, population dissipation from the states $\lvert 2_{20}\rangle$ and $\lvert 3_{01}\rangle$ cannot be fully decoupled. As a consequence,
state $\lvert 2_{20}\rangle$ and $\lvert 3_{01}\rangle$ apparently do not lead to a proper thermal behaviour at higher temperature.

\begin{figure}[htb!]
\centering
\includegraphics[width=0.45\textwidth,angle=-90]{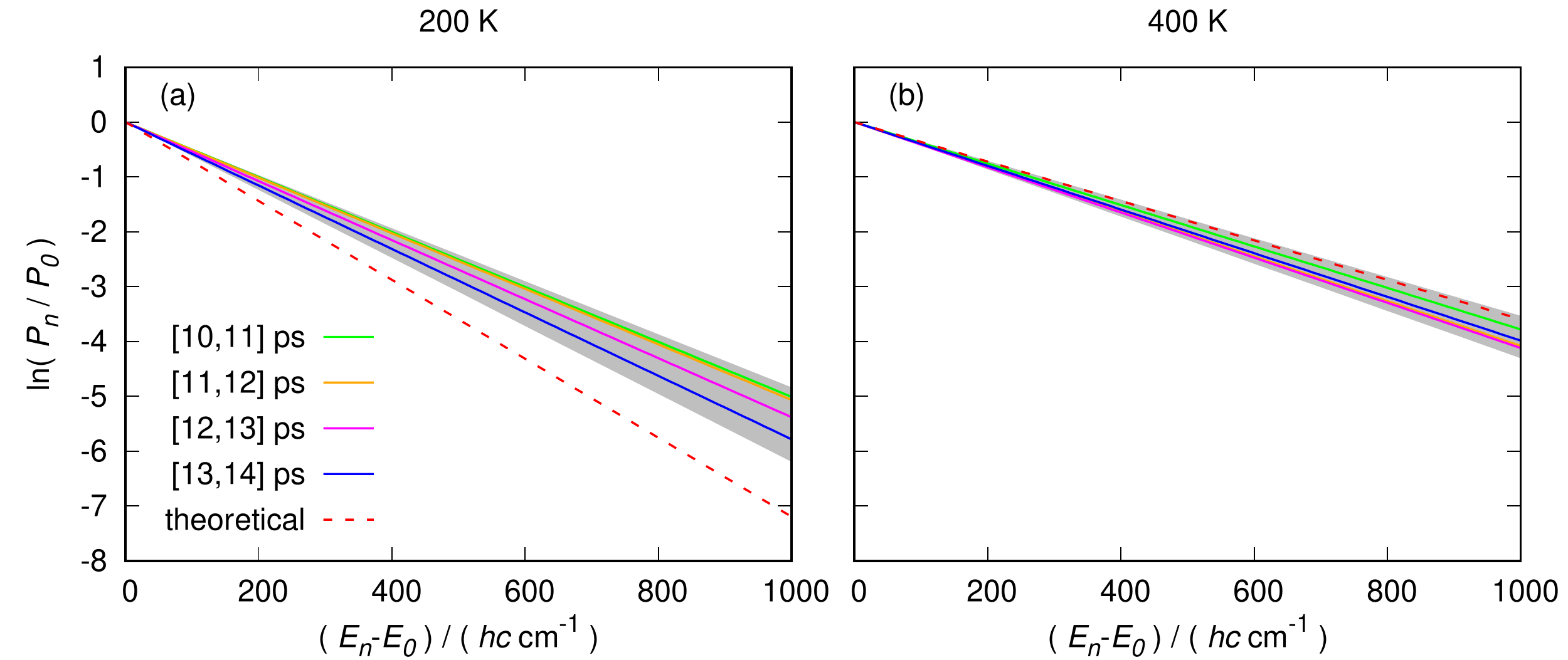}
\caption{Fit (solid line) of $\ln(P_n/P_0$) as a function of $E_n-E_0$ at bath temperatures 
(a) 200 K  and (b) 400 K evaluated from Figures \ref{pop_anhar_200K}(b) and \ref{pop_anhar_400K} at different time interval as indicated by the colour code key (valid for both panels). The gray shaded cone shows the maximum and minimum range among the errors related to all fits. The dashed red line is the theoretically expected behavior from the Boltzmann distribution at the respective bath temperatures.}
\label{temp_anhar_200K_400K}
\end{figure}

In order to better understand the thermalized nature of the equilibrated dynamics from the above populations, $\ln(P_n/P_0$) is calculated as a function of the energy difference to the ground state, $E_n-E_0$. For this study, dissipation dynamics is calculated with Eq. \eqref{relaxOP}, exclusively. The results are shown in Figures \ref{temp_anhar_200K_400K} (a) and (b) for 200 and 400 K, respectively. To evaluate the convergence of the dynamics with respect to the number of realizations of the stochastic process, the population $P_n$ of the $n^{\rm th}$ eigenstate is first averaged over a particular time interval at asymptotic times. Figure \ref{temp_anhar_200K_400K} depicts the linear regression of $\ln(P_n/P_0$) as a function of $E_n-E_0$ obtained for four different time intervals ([10,11], [11,12], [12,13] and [13,14] ps). All eigenstates with populations below 10$^{-6}$ are neglected for the fitting. While the average slopes are drawn as solid lines, the maximum and minimum range among the statistical errors obtained as the uncertainty on the slopes of all fits is shown by a shaded grey cone. The dashed red line is the theoretically calculated slope ($1/k_{\rm B} T$) from the Boltzmann distribution at 200 K (panel (a)) and 400 K (panel (b)). In both cases, the fits are well grouped and they reveal statistically significant trends. Figure \ref{temp_anhar_200K_400K}(a) shows that all fitted curves including their statistical error cone pass above the theoretical expected dashed red line. The system thus remains hyperthermal at 200 K for the anharmonic strong coupling potential despite the normal mode approach of Eq. \eqref{relaxOP}. This hyperthermal behaviour arises because the dissipation operators are only
 decoupled to second order. On the other hand, a slightly hypothermal behaviour is seen in Figure \ref{temp_anhar_200K_400K}(b) at a bath temperature of 400 K. This different behaviour compared to a system at 200 K is possibly due to the choice of the temperature-dependence for the coefficients $\mu(T)$ and $\nu(T)$ (see Eq. \eqref{coeffs1} and \eqref{coeffs2}), which is physically well-motivated~\cite{Gao:1998} but not strictly valid at all temperatures.~\cite{Ford:1999} We shall review this point later on in this work. In general, although the general trends in population for strongly coupled systems are in much better agreement with the ensemble temperatures, the detailed behaviour remains somewhat difficult to predict.

\subsection{Fermi resonance potential}\label{fermi}

\begin{figure}[tb!]
\centering
\includegraphics[width=0.42\textwidth,angle=-90]{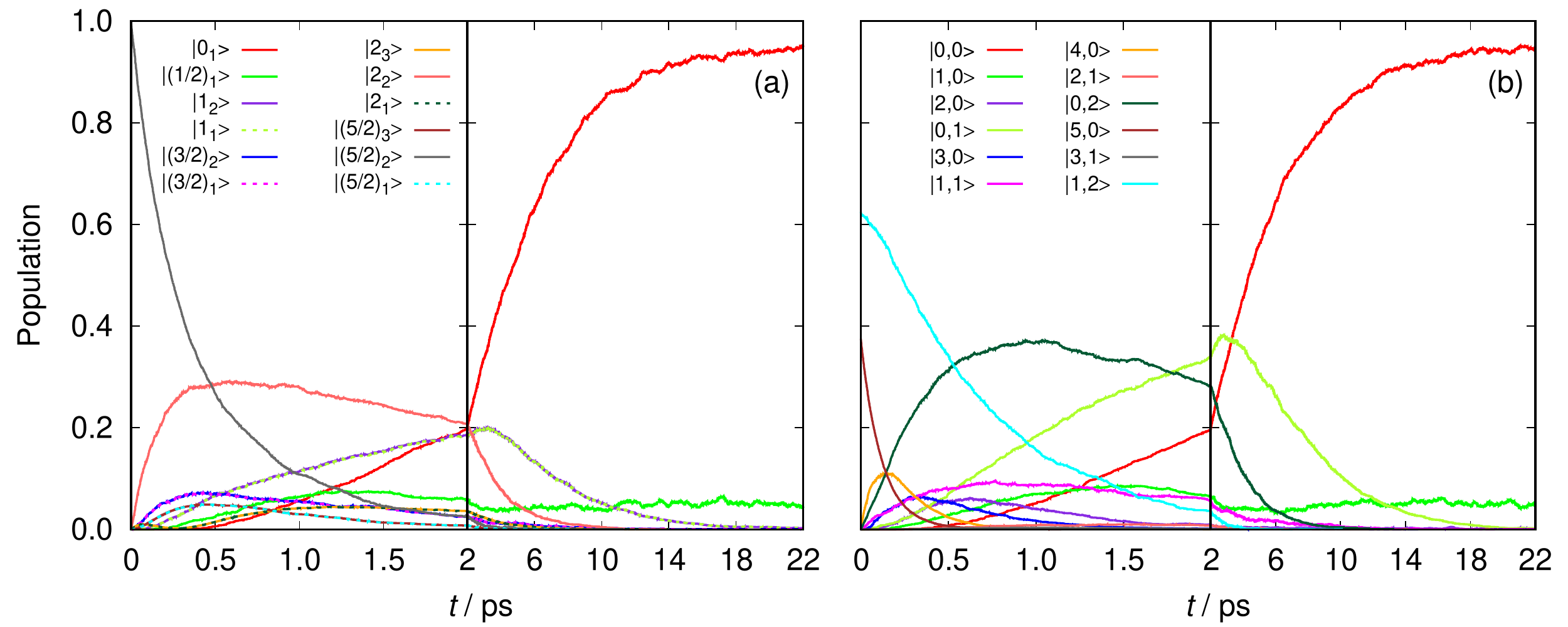}
\caption{Evolution of population of selected (a) eigenstates and (b) zeroth-order states for a two-dimensional harmonic Fermi resonance potential (see Eq. \eqref{eq:fermi_ham}) coupled to a bath at 200 K.
The initial wavefunction is chosen as an eigenstate $\lvert (5/2)_2 \rangle$ of the system hamiltonian (see Eq. \eqref{eq:fermi_ham}).
The relaxation parameters are $\gamma_Z^{-1}$ = 0.5 ps and $\gamma_R^{-1}$ = 2 ps and the coupling strength is $C$ = 0.1. Within a same polyad some eigenstates have almost equal population throughout the simulation period. Those states are \{$\lvert 1_2 \rangle$, $\lvert 1_1 \rangle$\}, \{$\lvert (3/2)_2 \rangle$, $\lvert (3/2)_1 \rangle$\}, \{$\lvert 2_3 \rangle$, $\lvert 2_1 \rangle$\}, and \{$\lvert (5/2)_3 \rangle$, $\lvert (5/2)_1 \rangle$\}.}
\label{pop_harmonic_Fermi_C_0p1_200K_eigen_zero_order}
\end{figure}

Figure \ref{pop_harmonic_Fermi_C_0p1_200K_eigen_zero_order} displays the population dynamics  for the harmonic Fermi resonance potential (see Eq. \eqref{eq:fermi_ham}) with a bath temperature of 200 K for weak intermode coupling ($C$ = 0.1). The left and right panels show the population dynamics of the eigenstates (see Figure \ref{pop_harmonic_Fermi_C_0p1_200K_eigen_zero_order}(a)) and zeroth-order states (see Figure \ref{pop_harmonic_Fermi_C_0p1_200K_eigen_zero_order}(b)) of the system hamiltonian (see Eq. \eqref{eq:fermi_ham}), respectively. The dissipation operator is represented by Eq. \eqref{relaxOP} in both cases. The initial wavefunction for both cases is chosen as eigenstate $\lvert (5/2)_2\rangle$ of the dissipation-free hamiltonian $\hat{H}_{\rm sys}$
(see Eq. \eqref{eq:fermi_ham}).
As a result the initial population is found completely in eigenstate $\lvert (5/2)_2 \rangle$ when reporting the population dynamics in the eigenstates basis (see Figure \ref{pop_harmonic_Fermi_C_0p1_200K_eigen_zero_order}(a)).
On the other hand, states $\lvert 1,2 \rangle$ (major with 63\% population) and  $\lvert 5,0 \rangle$ (minor with 37\% population) dominate the initial population in the zeroth-order basis (see Figure \ref{pop_harmonic_Fermi_C_0p1_200K_eigen_zero_order}(b)). This shows that the initial state $\lvert (5/2)_2 \rangle$ is a strong mixture of $\lvert 1,2\rangle$ and $\lvert 5,0 \rangle$, with almost no
contribution of $\lvert 3,1 \rangle$.
At early times (before 2ps), the dynamics in the eigenstate basis follows two paths.
One path is the sequential decay from one polyad to the next.
The population dynamics (see Fig.\,\ref{pop_harmonic_Fermi_C_0p1_200K_eigen_zero_order}(a)) suggests that this sequential path alternates between fast and slow population transfers.
We attribute this behaviour to the strong character mixing in the Fermi resonance. At each step, the relaxation is dominated either by its fast component ($Z$-mode, $\gamma_Z^{-1}=500$\,fs) or its slow component ($R$-mode, $\gamma_Z^{-1}=2$\,ps). 

The second path involves redistribution of energy within the $(5/2)$ polyad via a slow transition. Given that these states are eigenstates of the system, this slow transition must involve the bath. It is followed by a fast transition to the $\{\vert 2_3\rangle,\vert 2_1\rangle\}$ states, and a slow transition to the $(3/2)$ polyad. These two paths can be depicted as  (scheme I)\\

\begin{tikzpicture}
\node (522) at (0.0,0) { $\lvert (5/2)_2\rangle$};
\node (22)[right of = 522, xshift=1.9cm] { $\lvert 2_2\rangle$};
\node (32)[right of =22, xshift=1.9cm,rectangle, rounded corners,  align=center,draw]{$\lvert (3/2)_2\rangle$ \\ $\lvert (3/2)_1\rangle$};
\node (1)[right of =32, xshift=1.9cm,rectangle, rounded corners,  align=center,draw]{$\lvert 1_2\rangle$ \\ $\lvert 1_1\rangle$};
\node (121)[right of = 1, xshift=1.9cm] { $\lvert (1/2)_1\rangle$};
\node (01)[right of = 121, xshift=1.9cm] { $\lvert 0_1\rangle$};
\node (52)[below of =522, xshift=0cm,yshift=-1.5cm,rectangle, rounded corners,  align=center,draw]{$\lvert (5/2)_3\rangle$ \\ $\lvert (5/2)_1\rangle$};
\node (2)[right of =52, xshift=4.8cm,rectangle, rounded corners,  align=center,draw]{$\lvert 2_3\rangle$ \\ $\lvert 2_1\rangle$};
\draw [->] (522) --node[above] {F} (22);
\draw [->] (22) --node[above] {S} (32);
\draw [->] (32) --node[above] {F} (1);
\draw [->] (1) --node[above] {S} (121);
\draw [->] (121) --node[above] {F} (01);
\draw [->] (522) --node[left]{S} (52);
\draw [->] (52) --node[below] {F} (2);
\draw [->] (2) --node[right] {S} (32);
\node (nb1)[right of = 2, xshift=1.0cm,yshift=0.5cm,anchor=west] {F: Fast};
\node (nb2)[right of = 2, xshift=1.0cm,yshift=-0.1cm,anchor=west] {S: Slow};
\end{tikzpicture}

The energy relaxation behaviour can be alternatively understood with the help of zeroth-order states. Two relaxation paths can be identified from the two dominantly populated zeroth-order states (scheme II)\\

\begin{tikzpicture}\centering
\node (37percent) at (0.0,0) {minor (37\%)};
\node (50)[xshift=1cm, right of = 37percent] { $\lvert 5,0\rangle$};
\node (40)[right of = 50, xshift=2cm] { $\lvert 4,0\rangle$};
\node (30)[right of = 40, xshift=2cm] { $\lvert 3,0\rangle$};
\node (20)[right of = 30, xshift=2cm] { $\lvert 2,0\rangle$};
\node (10)[right of = 20, xshift=2cm] { $\lvert 1,0\rangle$};
\draw [->] (50) --node[above] {F (0.1 ps)} (40);
\draw [->] (40) --node[above] {F (0.125 ps)} (30);
\draw [->] (30) --node[above] {F (0.167 ps)} (20);
\draw [->] (20) --node[above] {F (0.25 ps)} (10);
\node (63percent)[below of = 37percent, yshift=-2cm] {major (63\%)};
\node (12)[right of = 63percent, xshift=1cm] { $\lvert 1,2\rangle$};
\node (02)[right of = 12, xshift=2cm] {$\lvert 0,2\rangle$};
\node (11)[right of = 12, xshift=5cm,yshift=-2cm] { $\lvert 1,1\rangle$};
\node (01)[right of = 02, xshift=5cm] {$\lvert 0,1\rangle$};
\node (00)[right of = 01, xshift=2cm] {$\lvert 0,0\rangle$};
\draw [->] (12) --node[above, sloped] {F (0.5 ps)} (02);
\draw [->] (12) --node[below, sloped] {S (1 ps)} (11);
\draw [->] (02) --node[above, xshift=-1.5cm] {S (1 ps)} (01);
\draw [->] (11) --node[below, sloped] {F (0.5 ps)} (01);
\draw [->] (01) --node[above] {S (2 ps)} (00);
\draw [->] (10) --node[left] {F (0.5 ps)} (00);
\draw [dashed, <->](50) --node[left] {IC (5.2 ps)}(12);
\draw [dashed, <->](40) --node[left] {IC (7.2 ps)}(02);
\draw [dashed, <->](30) --node[left, yshift=1cm] {IC (11.9 ps)}(11);
\draw [dashed, <->](20) --node[left] {IC (20.8 ps)}(01);
\node (nb1)[right of = 01, xshift=1.65cm,yshift=-1cm] {F: Fast};
\node (nb2)[right of = 01, xshift=1.6cm,yshift=-1.5cm] {S: Slow};
\node (nb3)[right of = 01, xshift=0.25cm,yshift=-2cm] {IC: Internal Conversion};
\end{tikzpicture}

\noindent{}The timescales given in parenthesis are estimated from the rate parameters $\gamma_{Z/R}$ in Eq.\,\eqref{coeffs1} using the harmonic scaling law.
The internal conversion (IC) timescales refer to population transfer within a polyad and are calculated from the energetic splitting within that polyad.
The early-time dynamics in Figure 4(b) reveals that the minor component, state $\lvert 5,0\rangle$, which is a highly excited state along the $Z$-mode relaxes sequentially and rapidly since the associate rate is faster.
The major channel starts with the fast $\lvert 1,2\rangle\to\lvert 0,2\rangle$ transition mediated by the loss of one quantum 
along the $Z$-mode due to relaxation via the bath.
Internal conversion between states $\lvert 0,2\rangle$ and $\lvert 4,0\rangle$ is possible because the two states are part of the same 
polyad. Consequently, the relaxation would continue along the fast sequential path along the $Z$-mode (upper row in scheme II). The alternative path involves relaxation of a quantum in the $R$-mode to reach the $\lvert 0,1\rangle$ state (lower row in scheme II).
It can also be reached via the $\lvert 1,1\rangle$ state in 
two steps, first by loosing a quantum in the slow $R$-mode and then in the fast $Z$-mode. 
The population dynamics reveals that these two sequential paths with alternating slow-fast relaxation towards state $\lvert 0,1\rangle$ are the dominant ones.

In the zeroth-order representation, population accumulates in the lowest-lying excited state along the $R$-mode. This is because relaxation along $R$ is slower.
In the eigenstate basis, population accumulates
        in the polyad, $1$.
It is the lowest-lying polyad that contains strong character of the $R$-mode.
The energy of the $Z$-mode is half that of the $R$-mode.
The state $\lvert (1/2)_1 \rangle $ is dominated by the $Z$-character, which undergoes faster relaxation than the $R$-mode.
                Thus, population transfer from the $1$ polyad to state $\vert (1/2)_1\rangle$ is the rate limiting step.

\begin{figure}[tb!]
\centering
\includegraphics[width=0.45\textwidth,angle=-90]{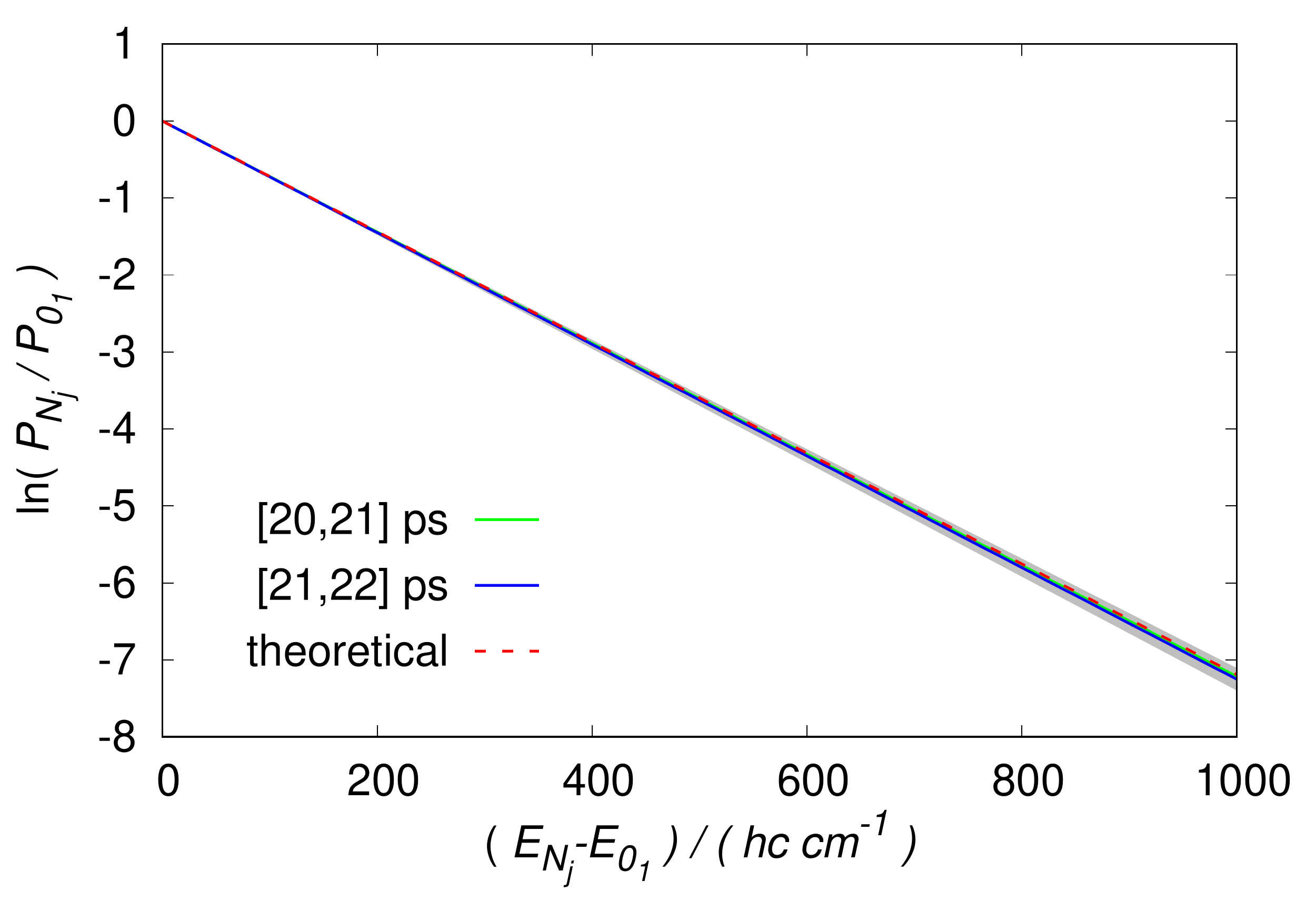}
\caption{Fit (solid line) of $\ln(P_{N_j}/P_{0_1})$ as a function of $E_{N_j}$ -$E_{0_1}$ for a bath temperature of 200 K evaluated from Figure \ref{pop_harmonic_Fermi_C_0p1_200K_eigen_zero_order}(a) at two different time intervals as indicated by the colour code. The gray shaded cone shows the maximum and minimum range among the errors related to all fits. The dashed red line is the theoretically expected behavior from the Boltzmann distribution at the bath temperature.}
\label{lnpnbyp0_with_error_harmonic_fermi_resonance_200K}
\end{figure}

By construction, populations for both eigenstates and zeroth-order states  (see Figure \ref{pop_harmonic_Fermi_C_0p1_200K_eigen_zero_order}(a) and  \ref{pop_harmonic_Fermi_C_0p1_200K_eigen_zero_order}(b)) reach constant equilibrium values at asymptotic times.
The  coefficients Eqs. \eqref{coeffs1} to \eqref{coeffs3} used in the dissipation operators are chosen to aim for a thermal equilibrium, at least in the case of
uncoupled harmonic oscillators. 
The decoupling approach proposed in this work yields thermal equilibrium also in the case of resonant coupling. To verify this, the population $P_{N_j}$ of the eigenstates $\lvert N_j \rangle$ in Figure \ref{pop_harmonic_Fermi_C_0p1_200K_eigen_zero_order}(a)
is used to fit $\ln(P_{N_j}/P_{0_1})$  as a function of energy $E_{N_j}$ -$E_{0_1}$.
Figure \ref{lnpnbyp0_with_error_harmonic_fermi_resonance_200K} depicts the result of the linear fits (shown by the colour coded solid lines) when the asymptotic populations are averaged over time intervals [20,21] and [21,22] ps. The gray shaded cone shows the maximum and minimum range among the associated statistical error related to the fitting. The colour coded lines essentially overlap, which indicates that the calculations are statistically converged. The theoretically expected dashed red line from a Boltzmann distribution at 200 K fits very well within the error of the fits.
A similarly good result can be obtained for strong intermode coupling ($C=0.5$).
This implies that the Fermi resonance model system
can be decoupled and thermalized very effectively using the 
approach based on normal modes in the Lindblad operator.

\begin{figure}[tb!]
\centering
\includegraphics[width=0.43\textwidth,angle=-90]{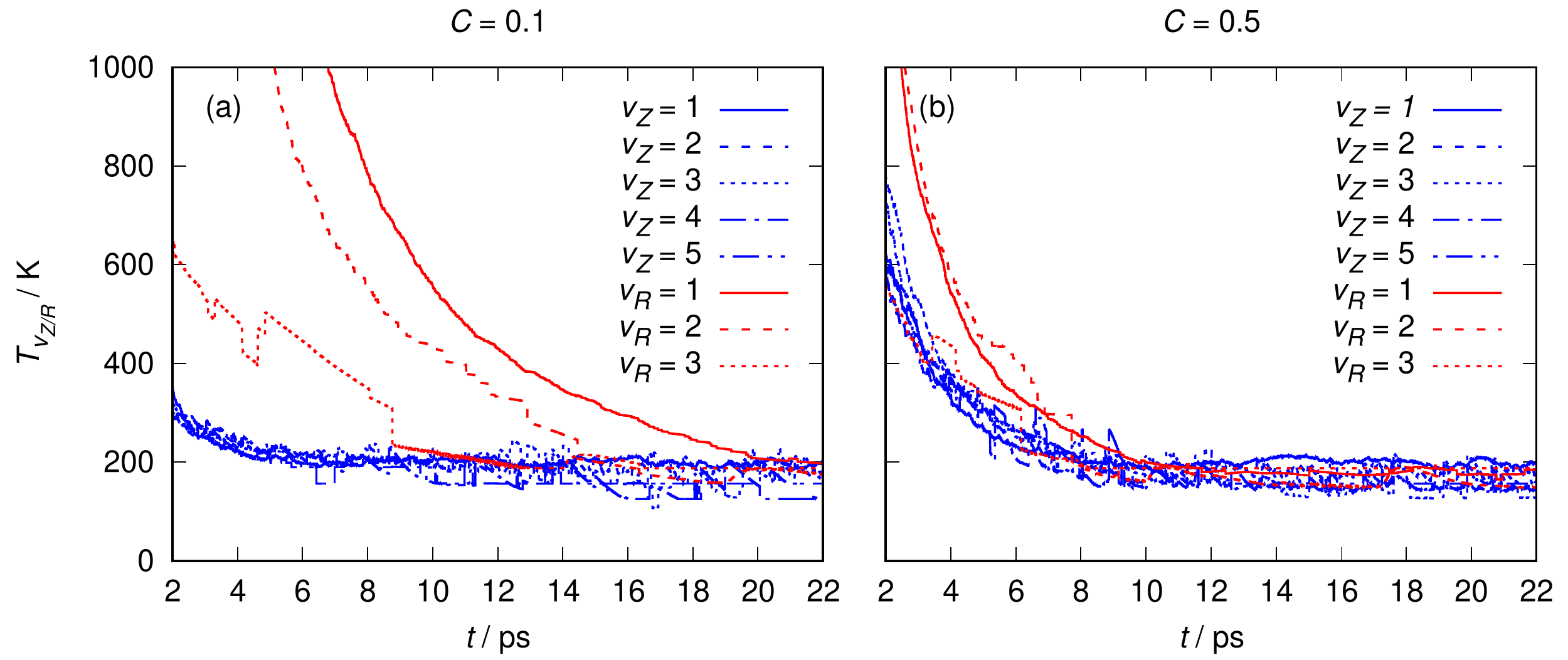}
\caption{Temperature $T_{v_{Z/R}}$ for different $v_Z$ and $v_R$ evaluated using Eq. \eqref{temperature} for Fermi resonance potential as a function of time for a bath at a temperature of  200 K. Panel (a) weak potential coupling, $C$ = 0.1. Panel (b) strong potential coupling, $C$ = 0.5.}
\label{reduced_temp_plot_harmonic_fermi_resonance_zeroth_order_C_0p1_200K}
\end{figure}

Despite the weak intermode potential coupling in the example discussed, the presence of the Fermi resonance leads to a strong degree of mode mixing in the model system.
This implies that both degrees of freedom should reach the same temperature, while this is not necessarily the case for coupled modes in general (see next section and Appendix).
Figure \ref{reduced_temp_plot_harmonic_fermi_resonance_zeroth_order_C_0p1_200K} depicts the time evolution of  temperatures calculated from the reduced populations of zeroth-order states along each mode.
These quantities are calculated from the dynamics reported in Figure \ref{pop_harmonic_Fermi_C_0p1_200K_eigen_zero_order}(b) as partial traces:
\begin{eqnarray}
P_{v_Z}^{({\rm r})}(t) &=& \sum_{v_R}P_{v_Z\; v_R} (t)\\
P_{v_R}^{({\rm r})}(t) &=& \sum_{v_Z}P_{v_Z\; v_R} (t) \, ,
\end{eqnarray}
where $P_{v_Z\; v_R} (t)$ is the population for state $\lvert v_Z, v_R\rangle$ at time $t$.
The state temperature $T_{v_{Z/R}}(t)$ is defined by inverting the Boltzmann distribution
\begin{eqnarray}
T_{v_{Z/R}}(t)= \frac{\epsilon_{v_{Z/R}} -\epsilon_0}{k_{\rm B}} \;\;\Bigg\{\ln \Bigg(\frac{P_0^{({\rm r})}(t)}{P_{v_{Z/R}}^{({\rm r})}(t)}\Bigg)\Bigg\}^{-1} \,.
\label{temperature}
\end{eqnarray}
Here, $\epsilon_{v_{Z/R}}$ is the energy of the $v_{Z/R}$th vibrational state of mode $Z/R$, respectively,
defined from the 1D potentials given in Eq.\,\eqref{eq:fermi_ham} by putting $C = 0$.
The temperatures calculated from Eq.\,\eqref{temperature} for the reduced populations of states $v_Z = \{1, 2, 3, 4, 5\}$ (blue lines) and $v_R = \{1, 2, 3\}$ (red lines)
are presented as a function of time in Figure \ref{reduced_temp_plot_harmonic_fermi_resonance_zeroth_order_C_0p1_200K} (a)  for $C=0.1$ and in Figure \ref{reduced_temp_plot_harmonic_fermi_resonance_zeroth_order_C_0p1_200K} (b) for $C=0.5$.

All states in both the $Z$- and $R$-modes start with different initial temperatures -- here depicted from 2 ps onward --
which are generally high compared to the bath temperature (200 K).
Eventually, they all reach the bath temperature.
The $Z$-mode has the fast relaxation rate in the model and 
all associated excited states rapidly reach the same temperature, following the same type of exponential decay.
The slower $R$-mode behaves differently. Since relaxation from a higher excited state is faster, the $v_R=3$ population reaches first the bath temperature, followed by the $v_R=2$ population. The temperature associated with the $v_R=1$ state remains hyperthermal for a longer time before also reaching the bath temperature.
The fact that all reduced populations of a given mode reach the same temperature is an indication that the system is found at thermal equilibrium (see also Figure \ref{lnpnbyp0_with_error_harmonic_fermi_resonance_200K}). 

Despite the weak intermode coupling stemming from the potential, the  character of the individual modes is completely lost in the eigenstates because of the exact resonant mixing of the zeroth-order states, independently of the size of the coupling strength as shown by the result
for stronger potential coupling ($C$ = 0.5), in Figure \ref{reduced_temp_plot_harmonic_fermi_resonance_zeroth_order_C_0p1_200K} (b).
Here again, all temperatures stabilise at the bath temperature 200 K, but the equilibration occurs even more rapidly due to stronger intermode coupling. Interestingly, the stronger intermode coupling leads to higher initial temperatures in the $Z$-mode while the $R$-mode is found at lower temperature compared to the weak coupling case. Stronger intermode coupling  also induces faster relaxation and faster intramolecular vibrational energy redistribution.
While the Fermi resonance condition implies that coupled modes loose their character in the composition of the system eigenstates, the individual modes reach the same thermal equilibrium state as in the later case.

The fact that the temperature of the two coupled modes is the same in Figure \ref{reduced_temp_plot_harmonic_fermi_resonance_zeroth_order_C_0p1_200K} is due to the Fermi resonance. It is related to the equivalence of time and state average (ergodic theorem) as a result of an entirely coherent quantum dynamics, was shown previously at the example of the wavepacket dynamics of the infrared CH chromophore in $\rm CHX_3$ compounds \cite{Marquardt:1986}.
 The Fermi resonance is mediated by the combination of two factors.
First, the zeroth order frequencies for the $Z$- and $R$-mode are commensurable, i.e., $\omega_R/\omega_Z$ = 2:1 in the present case.
Second, the  specific form of the coupling  chosen in the model hamiltonian is $\sim y_Ry_Z^2$ (see Eq. \eqref{eq:fermi_ham}). Such an operator can be expressed in terms of creation and annihilation operators,
where one quantum of vibration in the $R$-mode is created while two quanta of vibration in the $Z$-mode are destroyed, and vise versa.
These two conditions being met, strong intermode mixing becomes possible.

\subsection{Bilinearly coupled harmonic potentials}

\begin{figure}[b!]
\centering
\includegraphics[width=0.43\textwidth,angle=-90]{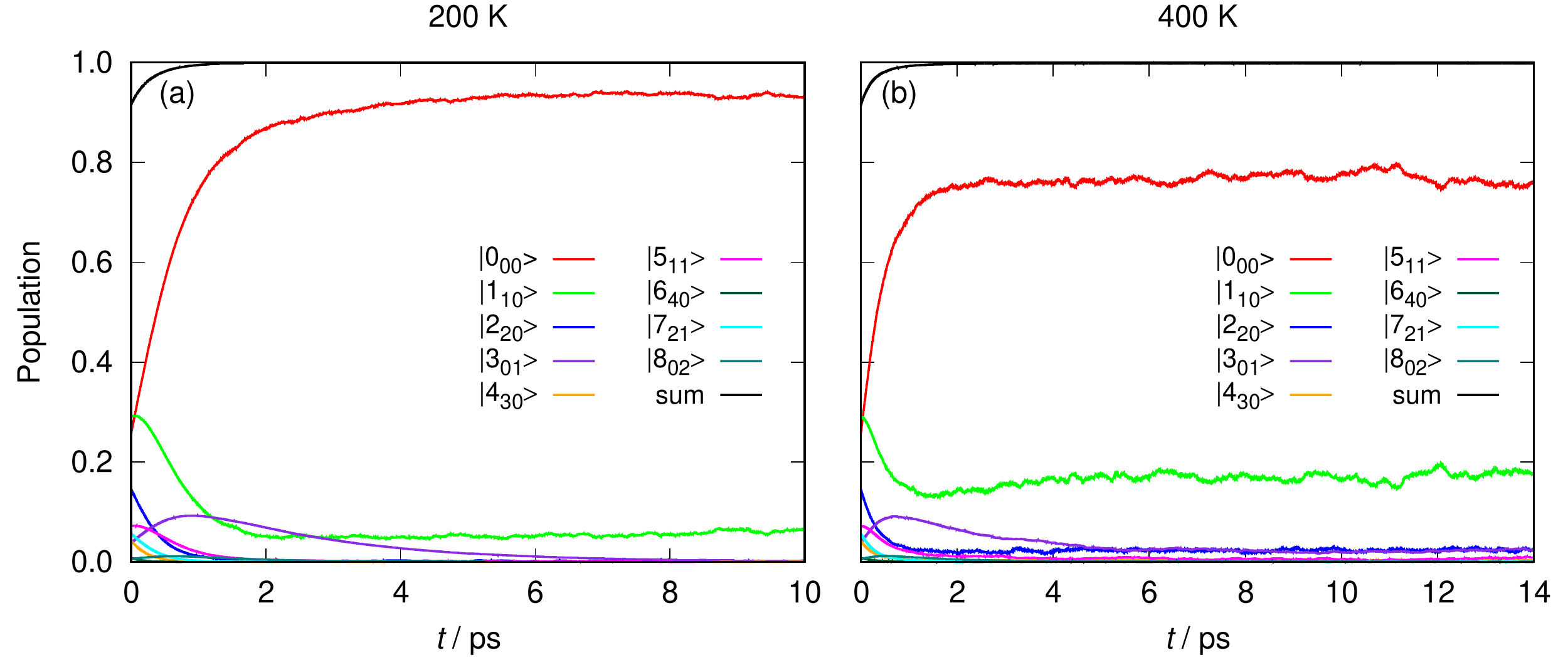}
\caption{Evolution of population of selected eigenstates within the two-dimensional model for O$_2$/Pt(111) at (a) 200 K and (b) 400 K for  the bilinearly coupled harmonic potentials (see Eq. \eqref{eq:hamonic_ham}). The initial wavefunction is chosen as a product of a  shifted Gaussian in $Z$ (center = 2.20 \AA, width = 0.039 \AA) 
and a Gaussian in $R$  (center = 1.37 \AA, width = 0.049 \AA).
The relaxation parameters are $\gamma_Z^{-1}$ = 0.5 ps and $\gamma_R^{-1}$ = 2 ps and the coupling strength is $C$ = 0.5.}
\label{fig:harmonic_200}
\end{figure}
Figure \ref{fig:harmonic_200} displays the population dynamics of eigenstates for the bilinearly coupled harmonic system (see Eq. \eqref{eq:hamonic_ham}) with strong coupling ($C$ = 0.5) using dissipation operators described by Eq. \eqref{relaxOP}.
The initial wavefunction is taken as a product of a shifted Gaussian in $Z$ and a Gaussian in $R$ centered at the equilibrium distance $R_{\rm e}$.  The simulations are carried out at bath temperatures of 200 K  and 400 K, and shown in panels (a) and (b), respectively.
The initial state predominantly populates the ground state ($\lvert 0_{00}\rangle$), as well as  excited states along the $Z$-mode ($\lvert 1_{10}\rangle $, $\lvert 2_{20}\rangle$ and $\lvert 4_{30}\rangle$) due to initial displacement in $Z$.
A few excited states in the $R$-mode ($\lvert 3_{01}\rangle $, $\lvert 5_{11}\rangle$ and $\lvert 7_{21}\rangle$) are also weakly populated due to the strong intermode coupling.
The figure shows the usual sequential exchanges of population among different states and modes, before finally reaching an equilibrium at asymptotic times. The asymptotic populations of the excited states are  higher for 400 K compared to 200 K, and the qualitative aspect of the equilibrium for the dominantly populated states is physically correct.

\begin{figure}[tb!]
\centering
\includegraphics[width=0.43\textwidth,angle=-90]{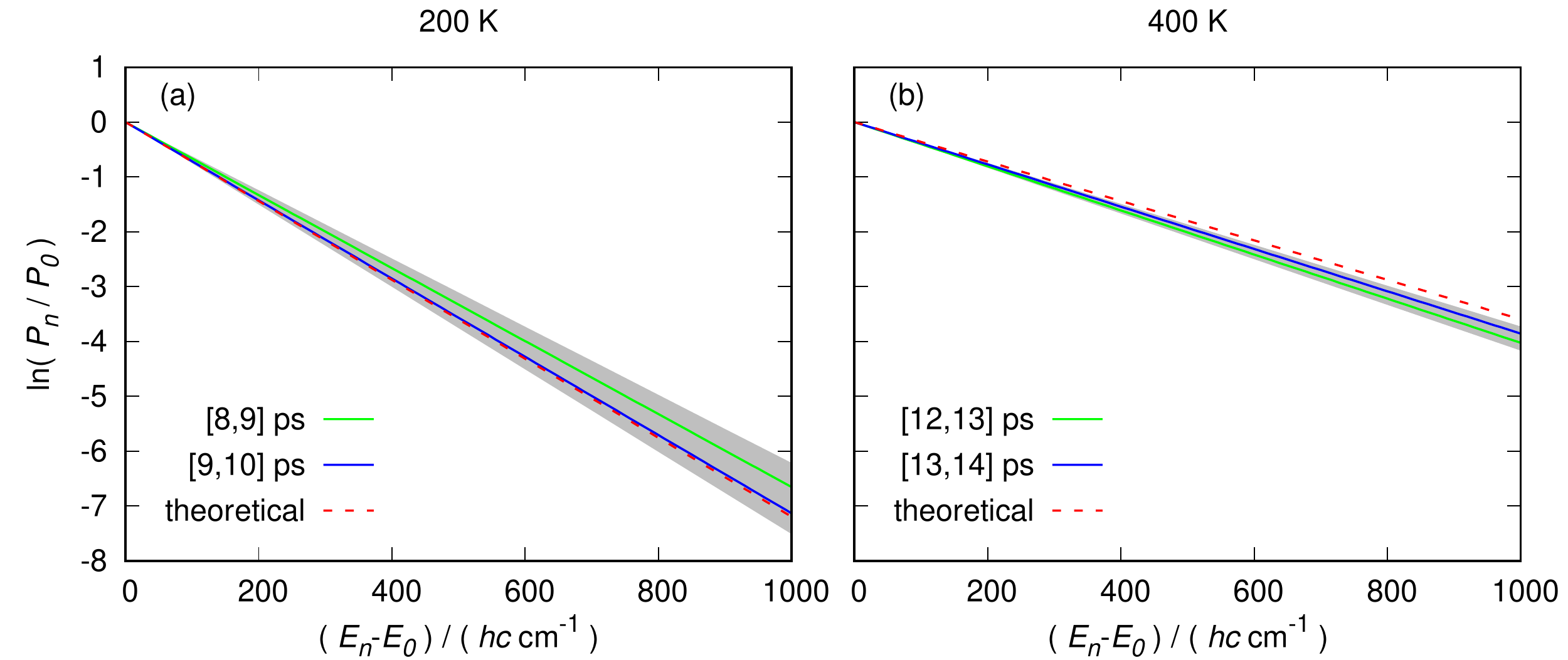}
\caption{Fit (solid line) of $\ln(P_n/P_0$) as a function of $E_n-E_0$ for bath temperatres (a) 200 K  and (b) 400 K evaluated from Figure \ref{fig:harmonic_200} at two different time intervals as indicated by
the colour code. The gray shaded cone shows the maximum and minimum range among the errors related to all fits. The dashed red line is the theoretically expected behavior from the Boltzmann distribution at the respective bath temperatures.}
\label{fig:temp_harmonic_200}
\end{figure}
In order to understand the nature of the equilibrium, the populations of the eigenstates from Figure \ref{fig:harmonic_200} are used to plot $\ln(P_{\rm n}/P_{0})$ as a function of energy $E_{\rm n}$ -$E_{\rm 0}$ using the method already discussed in Figures \ref{temp_anhar_200K_400K} and \ref{lnpnbyp0_with_error_harmonic_fermi_resonance_200K}. The results are plotted in Figure \ref{fig:temp_harmonic_200}(a) and (b) for 200 K and 400 K, respectively. Although the intermode coupling is strong the system reaches thermal equilibrium at 200 K within the error bars. Despite the improvement proposed by decoupling via normal mode analysis Eq. \eqref{relaxOP}, the population dynamics exhibits a hypothermal behaviour for a bath at a temperature of 400 K. Since the normal mode transformation should be exact for a bilinearly coupled harmonic system, this result is cause of concerns. The error introduced in the model at higher temperature likely stems from the definition of the coefficients used in the Lindblad operators. It is thus interesting to investigate the behaviour of the other type of generalized raising/lowering operators, defined by Eq.\,\eqref{thermal-relaxOP}.
\begin{figure}[tb!]
\centering
\includegraphics[width=0.43\textwidth,angle=-90]{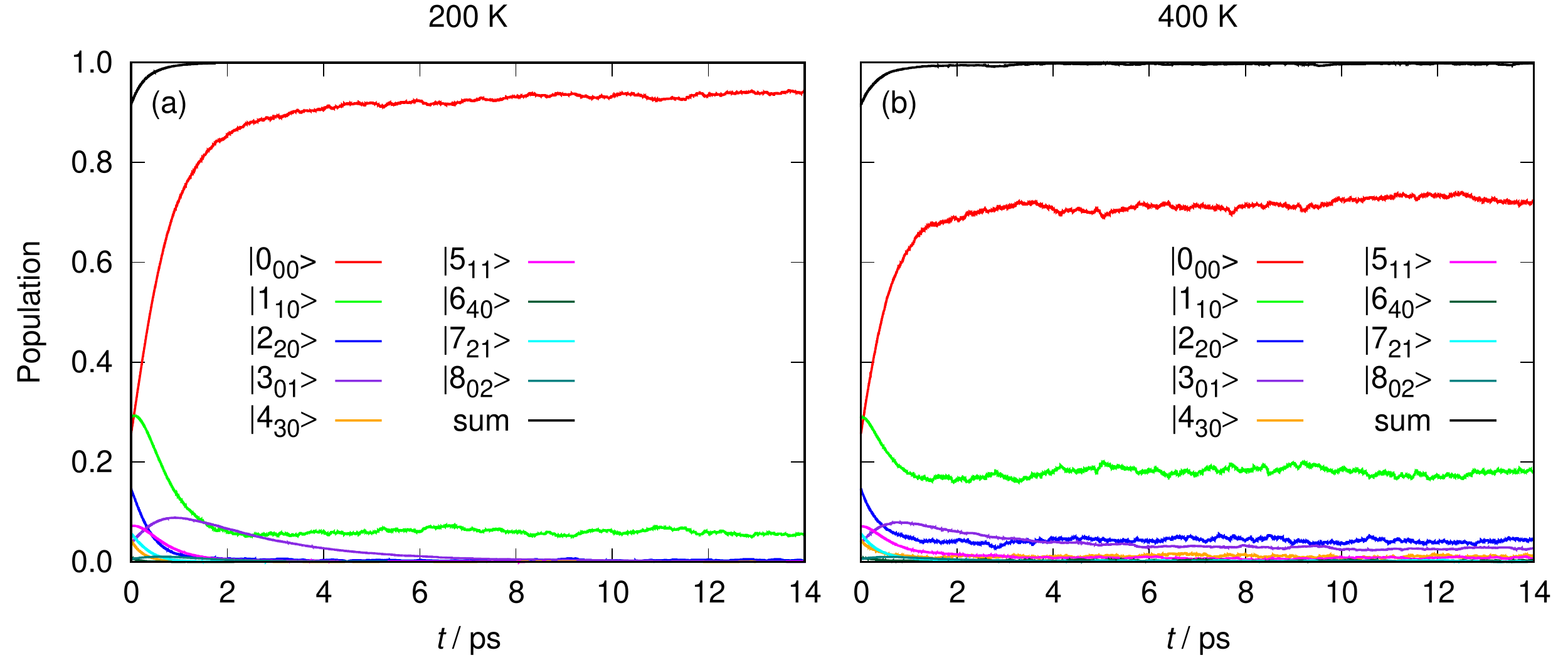}
\caption{Evolution of population of selected eigenstates basis within the two-dimensional model for O$_2$/Pt(111) at (a) 200 K and (b) 400 K for the bilinearly coupled harmonic potential (see Eq. \eqref{eq:hamonic_ham}) and modified dissipation operators based on Eqs. \eqref{BM-coff1} to \eqref{thermal-relaxOP}. The initial conditions, relaxation parameters and the coupling constant are same as Figuer \ref{fig:harmonic_200}.}
\label{fig:harmonic_200_RL}
\end{figure}

\begin{figure}[tb!]
\centering
\includegraphics[width=0.43\textwidth,angle=-90]{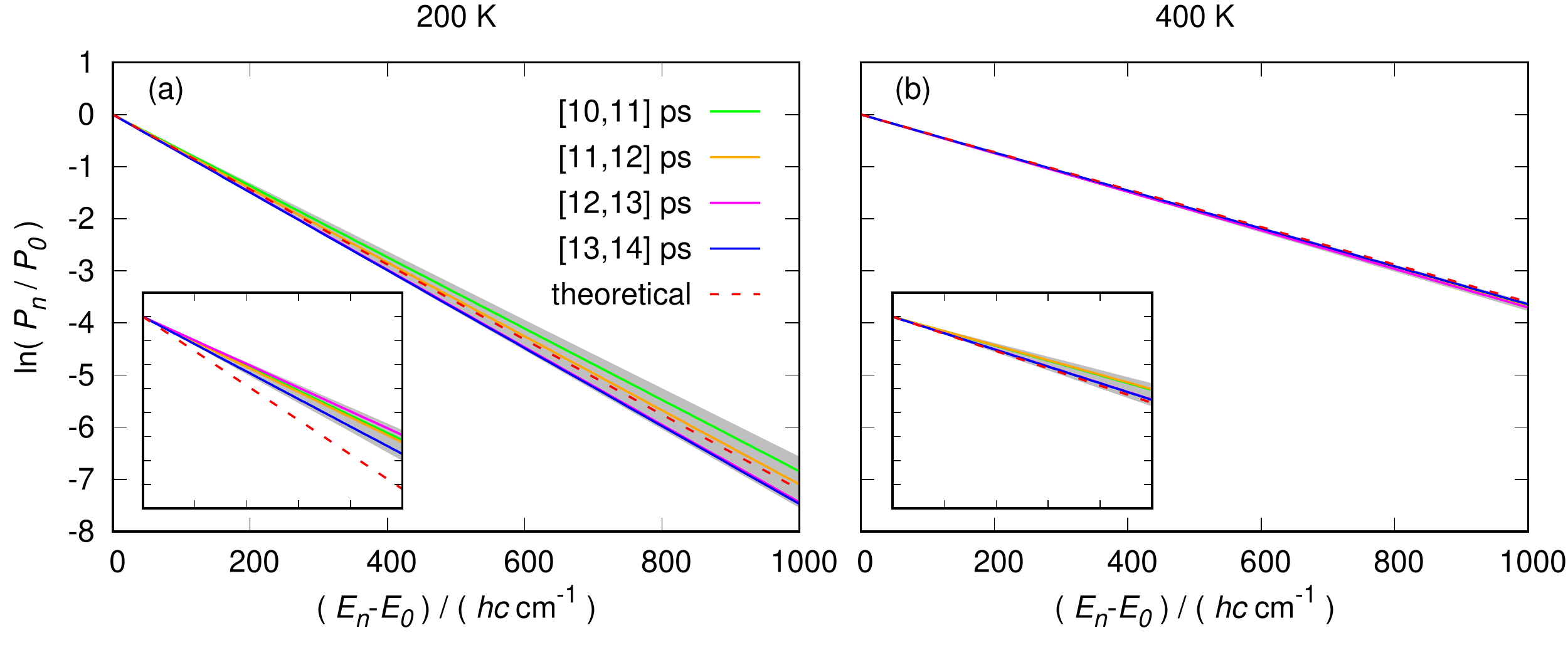}
\caption{Fit (solid line) of $\ln(P_n/P_0$) as a function of $E_n-E_0$ evaluated for the bilinearly coupled harmonic potentials as evaluated from Figure \ref{fig:harmonic_200_RL}(a) and \ref{fig:harmonic_200_RL} (b) at bath temperatures 200 K (a) and 400 K (b) and different time
intervals as indicated by the colour code key (valid for both panels).  The gray shaded cone shows the maximum and minimum range among the errors related to all fits. The dashed red line is the theoretically expected behavior from the Boltzmann distribution at the respective bath temperatures.  The insets show the equivalent results for the anharmonic strong coupling potential (see Eq. \eqref{eq:anharmonic_ham}).}
\label{fig:temp_harmonic_200_RL}
\end{figure}

The population dynamics of the eigenstates for the bilinearly coupled harmonic system is displayed in Figure \ref{fig:harmonic_200_RL}. 
The initial conditions are the same as described above. Panels (a) and (b) show the simulations at a bath temperature of 200 K and 400 K, respectively. 
At 200 K, the population dynamics is quantitatively the same as the one shown in Figure \ref{fig:harmonic_200}(a) using the original definition of the constants $\mu_k(T)$ and $\nu_k(T)$. The population of state $\lvert 3_{01}\rangle$ is visibly reduced at 400 K compared to Figure \ref{fig:harmonic_200} (b), with all remaining states bearing little to no population.
To provide a more quantitative picture of the system temperature, averaged populations are fitted according to the Boltzmann law as in the previous case studies.
Figure \ref{fig:temp_harmonic_200_RL}, panels (a) and (b) display the fit of $\ln(P_n/P_0$) as a function of $E_n-E_0$ at 200 K and 400 K, respectively.
 For the fit, the individual state populations are averaged over four different time intervals: [10,11], [11,12], [12,13] and [13,14] ps.
All eigenstates with population below 10$^{-6}$ are neglected for the fit.
The maximum and minimum range among the errors related to the fits (solid line) is shown by a shaded gray cone. The dashed red line is the theoretical line calculated from the Boltzmann distribution for 200 K and 400 K, respectively. In both cases, the theoretical dashed red line matches perfectly within the fitting error.

This confirms that the ensemble behaves as a thermal distribution at both the lower and higher temperature investigated here. At the lower temperature, the population of the higher excited states is much smaller and the slope is steeper.
Further, the numerical uncertainty in their determination is larger, such that the statistical error on the slope is larger than at the higher temperature. Finally, the earlier time interval, [10,11] ps (green line in Figure \eqref{fig:temp_harmonic_200_RL} (a)), yields a higher temperature than that of the bath, which is indicative that the system is not completely equilibrated at those early time intervals. Toward the end of the simulation (magenta and dark blue lines), the fitted temperature is slightly lower than the bath temperature. It is worth studying the effect of the new dissipation operators on the previously investigated system of coupled anharmonic potentials and to
compare the results with those obtained in section \ref{anharmonic}, in particular in Figure \ref{temp_anhar_200K_400K}. The plots obtained for the anharmonic strong coupling potential with the dissipation operators defined in Eq. (\eqref{thermal-relaxOP} are displayed in the insets of Figure \eqref{fig:temp_harmonic_200_RL}. The hyperthermal behaviour is still recognized at 200 K, but it is much smaller at 400 K. We attribute this to a more significant population of the higher excited states, which leads to smaller numerical errors in the populations and in the fits.

The hyperthermal behaviour arises because the dissipation operators defined in Eq. \eqref{thermal-relaxOP} do not depend on energy. The operators lead to the creation/annihilation of a well defined energy quantum. While such operators fit well the case of harmonic oscillators,
they fail in the case of anharmonic potentials.
Morse-type anharmonic oscillators for instance have levels with decreasing energetic spacing. Consequently, the raising/lowering dissipation operators fulfill the detailed balance condition only approximately at higher energies. This is likely the origin of the skewed population distribution.

\begin{figure}[tb!]
\centering
\includegraphics[width=0.5\textwidth,angle=-90]{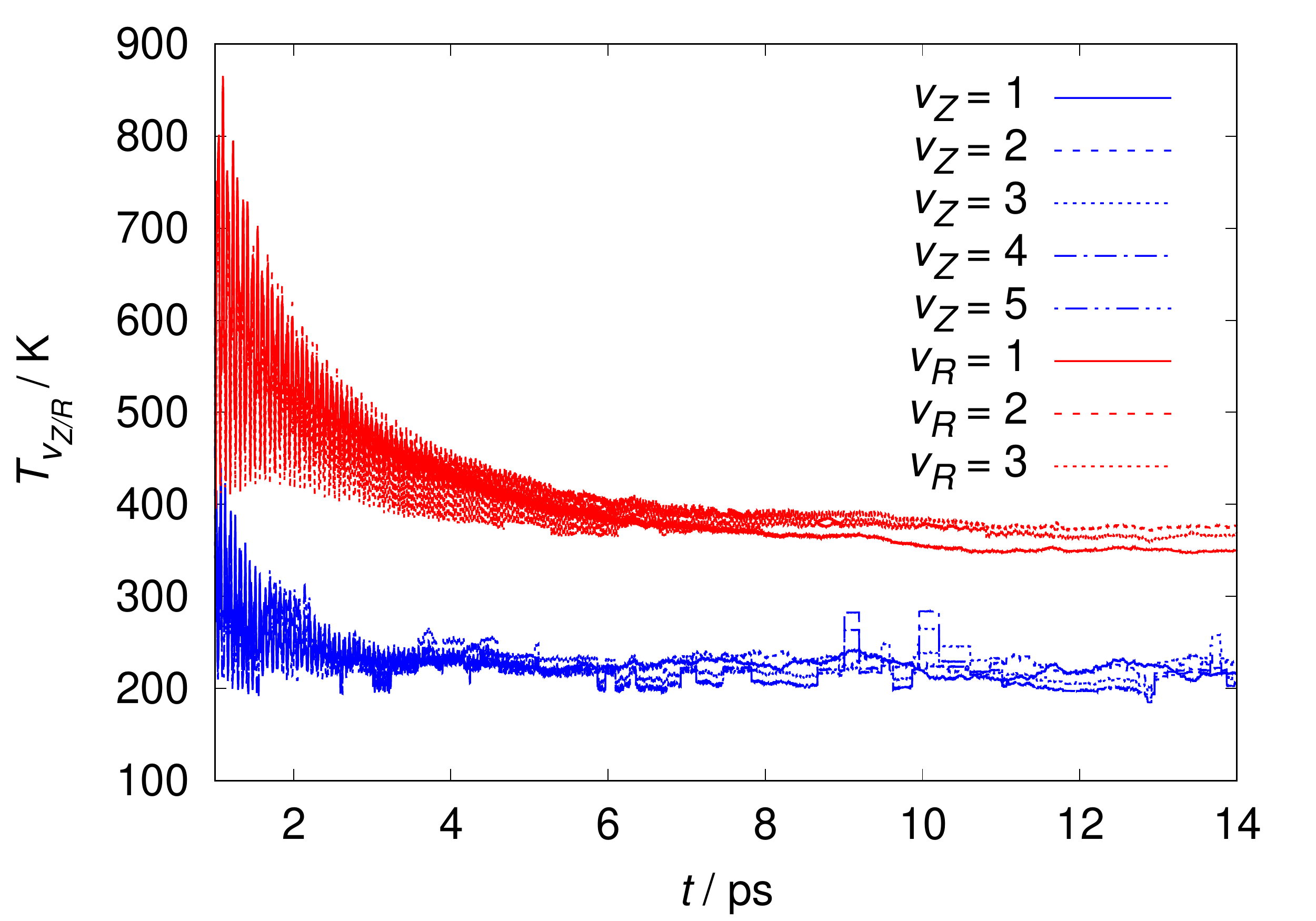}
\caption{Temperature $T_{v_{Z/R}}$ for different $v_Z$ and $v_R$ evaluated using Eq. \eqref{temperature} for the bilinearly coupled harmonic potential using modified dissipation operators based on Eqs. \eqref{BM-coff1} to \eqref{thermal-relaxOP} as a function of time for a bath at a temperature of  200 K.}
\label{fig:temp_from_reduced_pop_harmonic_200}
\end{figure}

In contrast to the Fermi resonance case discussed in the section \ref{fermi}, the two modes are not found at the same temperature for the bilinearly coupled harmonic potential even when the system has reached thermal equilibrium.
Figure \ref{fig:temp_from_reduced_pop_harmonic_200} depicts the time dependent mode temperatures calculated from Eq.\,\eqref{temperature} for the reduced populations of the lowest-lying states along the $Z$-mode ($v_Z = \{1, 2, 3, 4, 5\}$, blue lines)
and along the $R$-mode ($v_R = \{1, 2, 3\}$,  red lines) similarly to Figure \ref{reduced_temp_plot_harmonic_fermi_resonance_zeroth_order_C_0p1_200K}. While the temperatures for a given mode converge to the same value, the temperatures of the two modes are different. This shows that the modes belong to different thermal ensembles: the temperature of the low-frequency $Z$-mode is close to the bath temperature of 200 K, but the high-frequency $R$-mode has a temperature of  about 370 K. 

    A rationale for the asymptotic stabilisation of the modes at a temperature different from the bath and from the other modes is given in the Appendix, where it is shown that a thermalized ensemble of multi-dimensional eigenstates does not necessarily imply thermalization of individual modes of the zeroth-order basis. This finding is in stark contrast with the Fermi resonance, where strong mode mixing and thermalization at a single temperature was observed even for weak intermode coupling. In the present case, the strong intermode coupling does not lead to strong mode mixing. Rather, rapid coherent population transfer between the two modes takes place at early times, as observed from the fast oscillations of the individual state temperatures in Figure \ref{fig:temp_from_reduced_pop_harmonic_200}. 
At longer times, these coherent oscillations disappear and differential thermalization occurs. This is indicative of a very different physical origin of the intermode coupling. The Fermi resonance redistributes population among essentially isoenergetic states of different
 modes.

\section{Conclusion} \label{conclusion}
In conclusion, we have used the Monte Carlo wave packet method implemented in the Heidelberg MCTDH package to study dissipative dynamics in 2D model systems subject to strong intermode coupling. 
In our previous work,~\cite{Mandal2022sMCTDH} we have shown that under certain circumstances the dissipation operators constructed from the system coordinates can lead to a hyperthermal behaviour of modes due to intermode coupling. To alleviate this issue, we have investigated the potential of a linear transformation of the dissipation operators from the generalized coordinates describing the system to their normal mode representation, which reduces intermode coupling up to the second order. We have first studied the equilibration dynamics for three different model potentials in the strong coupling regime.
For the anharmonic strong coupling potential, although the use of normal modes improves the thermal behaviour at the asymptotic times compared to generalized coordinates, the overall trend from the Boltzmann distribution of states remains hyperthermal for a bath at 200 K temperature. This is because the transformation to the normal modes can not fully eliminate the intermode coupling due to the anharmonicity.
On the other hand, the population dynamics at 400 K shows a hypothermal behaviour. This is partly due to the temperature-dependent constants related to dissipation operators which do not strictly enforce thermal detailed balance at all temperatures.

In the case of the coupled harmonic Fermi resonance potential, the system achieves the expected thermal distribution. The Fermi resonance also mixes both modes strongly, even for weak coupling, and the individual mode characteristics are completely lost. As a consequence, both modes are observed to reach the same temperature at equilibrium.

Finally, we have studied the thermalization dynamics for a bimodal harmonic potential with strong bilinear coupling, where the modes are perfectly decoupled in the normal mode representation. Although the population dynamics at 200 K reaches the expected thermal distribution, a hypothermal behaviour is observed at 400 K, which stems in part from the  definition of the temperature-dependent constants in the dissipation operator.

An alternative form of generalized raising/lowering operators allowed us to enforce exactly the correct thermal distribution in the case of the bilinearly coupled harmonic potential. Despite the strong intramolecular coupling for the harmonic case, the two modes reach different temperatures at equilibrium because of the specific form of the intermode coupling. This is in contrast with the harmonic Fermi resonance case, where strong mode mixing dominates the thermalization dynamics of the individual modes due to exact commensurable frequencies and the specific form of  the leading coupling term. These new dissipation operators also perform well for anharmonic potentials, in particular at higher bath temperatures where higher excited states are more significantly populated. The small hyperthermal behaviour for the anharmonic case is related to the form of dissipation operators adopted in this work, which are not energy dependent. Finding energy dependent dissipation operators to get exact thermalization even for anharmonic cases is of future research interest.

\begin{acknowledgments}
  We acknowledge financial support from ANR through the QDDA project as
  well as from Universit\'e de Strasbourg, Universit\'e de Lorraine,
  Universit\'e de Paris-Saclay, and CNRS for computational resources. The authors also thank the CNRS International Research Network (IRN) "MCTDH" for funding support.
\end{acknowledgments}


\begin{thebibliography}{34}%
\makeatletter
\providecommand \@ifxundefined [1]{%
 \@ifx{#1\undefined}
}%
\providecommand \@ifnum [1]{%
 \ifnum #1\expandafter \@firstoftwo
 \else \expandafter \@secondoftwo
 \fi
}%
\providecommand \@ifx [1]{%
 \ifx #1\expandafter \@firstoftwo
 \else \expandafter \@secondoftwo
 \fi
}%
\providecommand \natexlab [1]{#1}%
\providecommand \enquote  [1]{``#1''}%
\providecommand \bibnamefont  [1]{#1}%
\providecommand \bibfnamefont [1]{#1}%
\providecommand \citenamefont [1]{#1}%
\providecommand \href@noop [0]{\@secondoftwo}%
\providecommand \href [0]{\begingroup \@sanitize@url \@href}%
\providecommand \@href[1]{\@@startlink{#1}\@@href}%
\providecommand \@@href[1]{\endgroup#1\@@endlink}%
\providecommand \@sanitize@url [0]{\catcode `\\12\catcode `\$12\catcode
  `\&12\catcode `\#12\catcode `\^12\catcode `\_12\catcode `\%12\relax}%
\providecommand \@@startlink[1]{}%
\providecommand \@@endlink[0]{}%
\providecommand \url  [0]{\begingroup\@sanitize@url \@url }%
\providecommand \@url [1]{\endgroup\@href {#1}{\urlprefix }}%
\providecommand \urlprefix  [0]{URL }%
\providecommand \Eprint [0]{\href }%
\providecommand \doibase [0]{http://dx.doi.org/}%
\providecommand \selectlanguage [0]{\@gobble}%
\providecommand \bibinfo  [0]{\@secondoftwo}%
\providecommand \bibfield  [0]{\@secondoftwo}%
\providecommand \translation [1]{[#1]}%
\providecommand \BibitemOpen [0]{}%
\providecommand \bibitemStop [0]{}%
\providecommand \bibitemNoStop [0]{.\EOS\space}%
\providecommand \EOS [0]{\spacefactor3000\relax}%
\providecommand \BibitemShut  [1]{\csname bibitem#1\endcsname}%
\let\auto@bib@innerbib\@empty
\bibitem [{\citenamefont {Breuer}\ and\ \citenamefont
  {Petruccione}(2006)}]{breuerpetrobook}%
  \BibitemOpen
  \bibfield  {author} {\bibinfo {author} {\bibfnamefont {H.-P.}\ \bibnamefont
  {Breuer}}\ and\ \bibinfo {author} {\bibfnamefont {F.}~\bibnamefont
  {Petruccione}},\ }\href@noop {} {\emph {\bibinfo {title} {The Theory of Open
  Quantum Systems}}}\ (\bibinfo  {publisher} {Oxford University Press,
  Oxford},\ \bibinfo {year} {2006})\BibitemShut {NoStop}%
\bibitem [{\citenamefont {Nitzan}(2006)}]{nitzan2006chemical}%
  \BibitemOpen
  \bibfield  {author} {\bibinfo {author} {\bibfnamefont {A.}~\bibnamefont
  {Nitzan}},\ }\href@noop {} {\emph {\bibinfo {title} {Chemical dynamics in
  condensed phases: relaxation, transfer and reactions in condensed molecular
  systems}}}\ (\bibinfo  {publisher} {Oxford university press},\ \bibinfo
  {year} {2006})\BibitemShut {NoStop}%
\bibitem [{\citenamefont {Blum}(2012)}]{blum2012density}%
  \BibitemOpen
  \bibfield  {author} {\bibinfo {author} {\bibfnamefont {K.}~\bibnamefont
  {Blum}},\ }\href@noop {} {\emph {\bibinfo {title} {Density matrix theory and
  applications}}},\ Vol.~\bibinfo {volume} {64}\ (\bibinfo  {publisher}
  {Springer Science \& Business Media},\ \bibinfo {year} {2012})\BibitemShut
  {NoStop}%
\bibitem [{\citenamefont {Mandal}\ \emph {et~al.}(2022)\citenamefont {Mandal},
  \citenamefont {Gatti}, \citenamefont {Bindech}, \citenamefont {Marquardt},\
  and\ \citenamefont {Tremblay}}]{Mandal2022sMCTDH}%
  \BibitemOpen
  \bibfield  {author} {\bibinfo {author} {\bibfnamefont {S.}~\bibnamefont
  {Mandal}}, \bibinfo {author} {\bibfnamefont {F.}~\bibnamefont {Gatti}},
  \bibinfo {author} {\bibfnamefont {O.}~\bibnamefont {Bindech}}, \bibinfo
  {author} {\bibfnamefont {R.}~\bibnamefont {Marquardt}}, \ and\ \bibinfo
  {author} {\bibfnamefont {J.-C.}\ \bibnamefont {Tremblay}},\ }\bibfield
  {title} {\enquote {\bibinfo {title} {Multidimensional stochastic dissipative
  quantum dynamics using a {L}indblad operator},}\ }\href@noop {} {\bibfield
  {journal} {\bibinfo  {journal} {J. Chem. Phys.}\ }\textbf {\bibinfo {volume}
  {156}},\ \bibinfo {pages} {094109} (\bibinfo {year} {2022})}\BibitemShut
  {NoStop}%
\bibitem [{\citenamefont {Gao}(1997)}]{Gao:1997}%
  \BibitemOpen
  \bibfield  {author} {\bibinfo {author} {\bibfnamefont {S.}~\bibnamefont
  {Gao}},\ }\bibfield  {title} {\enquote {\bibinfo {title} {Dissipative
  {Q}uantum {D}ynamics with a {L}indblad {F}unctional},}\ }\href {\doibase
  10.1103/PhysRevLett.79.3101} {\bibfield  {journal} {\bibinfo  {journal}
  {Phys. Rev. Lett.}\ }\textbf {\bibinfo {volume} {79}},\ \bibinfo {pages}
  {3101--3104} (\bibinfo {year} {1997})}\BibitemShut {NoStop}%
\bibitem [{\citenamefont {Gao}(1998)}]{Gao:1998}%
  \BibitemOpen
  \bibfield  {author} {\bibinfo {author} {\bibfnamefont {S.}~\bibnamefont
  {Gao}},\ }\bibfield  {title} {\enquote {\bibinfo {title} {{Lindblad approach
  to quantum dynamics of open systems}},}\ }\href {\doibase
  10.1103/PhysRevB.57.4509} {\bibfield  {journal} {\bibinfo  {journal} {Phys.
  Rev. B}\ }\textbf {\bibinfo {volume} {57}},\ \bibinfo {pages} {4509--4517}
  (\bibinfo {year} {1998})}\BibitemShut {NoStop}%
\bibitem [{\citenamefont {Nest}\ and\ \citenamefont {Meyer}(2003)}]{nes03:24}%
  \BibitemOpen
  \bibfield  {author} {\bibinfo {author} {\bibfnamefont {M.}~\bibnamefont
  {Nest}}\ and\ \bibinfo {author} {\bibfnamefont {H.-D.}\ \bibnamefont
  {Meyer}},\ }\bibfield  {title} {\enquote {\bibinfo {title} {Dissipative
  quantum dynamics of anharmonic oscillators with the {M}ulti-{C}onfiguration
  {T}ime-{D}ependent {H}artree ({MCTDH}) {M}ethod},}\ }\href@noop {} {\bibfield
   {journal} {\bibinfo  {journal} {J.~Chem.\ Phys.}\ }\textbf {\bibinfo
  {volume} {119}},\ \bibinfo {pages} {24} (\bibinfo {year} {2003})}\BibitemShut
  {NoStop}%
\bibitem [{\citenamefont {Nest}\ and\ \citenamefont
  {Saalfrank}(2000)}]{nest2000open}%
  \BibitemOpen
  \bibfield  {author} {\bibinfo {author} {\bibfnamefont {M.}~\bibnamefont
  {Nest}}\ and\ \bibinfo {author} {\bibfnamefont {P.}~\bibnamefont
  {Saalfrank}},\ }\bibfield  {title} {\enquote {\bibinfo {title} {Open-system
  quantum dynamics for gas-surface scattering: Nonlinear dissipation and mapped
  fourier grid methods},}\ }\href@noop {} {\bibfield  {journal} {\bibinfo
  {journal} {J. Chem. Phys.}\ }\textbf {\bibinfo {volume} {113}},\ \bibinfo
  {pages} {8753--8761} (\bibinfo {year} {2000})}\BibitemShut {NoStop}%
\bibitem [{\citenamefont {Nest}\ and\ \citenamefont
  {Meyer}(2002)}]{nest2002improving}%
  \BibitemOpen
  \bibfield  {author} {\bibinfo {author} {\bibfnamefont {M.}~\bibnamefont
  {Nest}}\ and\ \bibinfo {author} {\bibfnamefont {H.-D.}\ \bibnamefont
  {Meyer}},\ }\bibfield  {title} {\enquote {\bibinfo {title} {Improving the
  mapping mechanism of the mapped {F}ourier method},}\ }\href@noop {}
  {\bibfield  {journal} {\bibinfo  {journal} {Chem. Phys. Lett.}\ }\textbf
  {\bibinfo {volume} {352}},\ \bibinfo {pages} {486--490} (\bibinfo {year}
  {2002})}\BibitemShut {NoStop}%
\bibitem [{\citenamefont {Pesce}\ and\ \citenamefont
  {Saalfrank}(1997)}]{pesce1997free}%
  \BibitemOpen
  \bibfield  {author} {\bibinfo {author} {\bibfnamefont {L.}~\bibnamefont
  {Pesce}}\ and\ \bibinfo {author} {\bibfnamefont {P.}~\bibnamefont
  {Saalfrank}},\ }\bibfield  {title} {\enquote {\bibinfo {title} {{`Free'}
  nuclear density propagation in two dimensions the coupled-channel density
  matrix method and its application to inelastic molecule-surface
  scattering},}\ }\href@noop {} {\bibfield  {journal} {\bibinfo  {journal}
  {Chem. Phys.}\ }\textbf {\bibinfo {volume} {219}},\ \bibinfo {pages} {43--55}
  (\bibinfo {year} {1997})}\BibitemShut {NoStop}%
\bibitem [{\citenamefont {Pesce}\ and\ \citenamefont
  {Saalfrank}(1998)}]{pesce1998coupled}%
  \BibitemOpen
  \bibfield  {author} {\bibinfo {author} {\bibfnamefont {L.}~\bibnamefont
  {Pesce}}\ and\ \bibinfo {author} {\bibfnamefont {P.}~\bibnamefont
  {Saalfrank}},\ }\bibfield  {title} {\enquote {\bibinfo {title} {The coupled
  channel density matrix method for open quantum systems: Formulation and
  application to the vibrational relaxation of molecules scattering from
  nonrigid surfaces},}\ }\href@noop {} {\bibfield  {journal} {\bibinfo
  {journal} {J. Chem. Phys.}\ }\textbf {\bibinfo {volume} {108}},\ \bibinfo
  {pages} {3045--3056} (\bibinfo {year} {1998})}\BibitemShut {NoStop}%
\bibitem [{\citenamefont {Pesce}\ \emph {et~al.}(1998)\citenamefont {Pesce},
  \citenamefont {Gerdts}, \citenamefont {Manthe},\ and\ \citenamefont
  {Saalfrank}}]{pesce1998variational}%
  \BibitemOpen
  \bibfield  {author} {\bibinfo {author} {\bibfnamefont {L.}~\bibnamefont
  {Pesce}}, \bibinfo {author} {\bibfnamefont {T.}~\bibnamefont {Gerdts}},
  \bibinfo {author} {\bibfnamefont {U.}~\bibnamefont {Manthe}}, \ and\ \bibinfo
  {author} {\bibfnamefont {P.}~\bibnamefont {Saalfrank}},\ }\bibfield  {title}
  {\enquote {\bibinfo {title} {Variational wave packet method for dissipative
  photodesorption problems},}\ }\href@noop {} {\bibfield  {journal} {\bibinfo
  {journal} {Chem. Phys. Lett.}\ }\textbf {\bibinfo {volume} {288}},\ \bibinfo
  {pages} {383--390} (\bibinfo {year} {1998})}\BibitemShut {NoStop}%
\bibitem [{\citenamefont {Saalfrank}(2006)}]{chemrev}%
  \BibitemOpen
  \bibfield  {author} {\bibinfo {author} {\bibfnamefont {P.}~\bibnamefont
  {Saalfrank}},\ }\bibfield  {title} {\enquote {\bibinfo {title} {Quantum
  dynamical approach to ultrafast molecular desorption from surfaces},}\
  }\href@noop {} {\bibfield  {journal} {\bibinfo  {journal} {Chem. Rev.}\
  }\textbf {\bibinfo {volume} {106}},\ \bibinfo {pages} {4116} (\bibinfo {year}
  {2006})}\BibitemShut {NoStop}%
\bibitem [{\citenamefont {Tremblay}, \citenamefont {Beyvers},\ and\
  \citenamefont {Saalfrank}(2008)}]{tremblay2008selective}%
  \BibitemOpen
  \bibfield  {author} {\bibinfo {author} {\bibfnamefont {J.~C.}\ \bibnamefont
  {Tremblay}}, \bibinfo {author} {\bibfnamefont {S.}~\bibnamefont {Beyvers}}, \
  and\ \bibinfo {author} {\bibfnamefont {P.}~\bibnamefont {Saalfrank}},\
  }\bibfield  {title} {\enquote {\bibinfo {title} {Selective excitation of
  coupled co vibrations on a dissipative {C}u (100) surface by shaped infrared
  laser pulses},}\ }\href@noop {} {\bibfield  {journal} {\bibinfo  {journal}
  {J. Chem. Phys.}\ }\textbf {\bibinfo {volume} {128}},\ \bibinfo {pages}
  {194709} (\bibinfo {year} {2008})}\BibitemShut {NoStop}%
\bibitem [{\citenamefont {Tremblay}\ and\ \citenamefont
  {Saalfrank}(2009)}]{tremblay2009selective}%
  \BibitemOpen
  \bibfield  {author} {\bibinfo {author} {\bibfnamefont {J.~C.}\ \bibnamefont
  {Tremblay}}\ and\ \bibinfo {author} {\bibfnamefont {P.}~\bibnamefont
  {Saalfrank}},\ }\bibfield  {title} {\enquote {\bibinfo {title} {Selective
  subsurface absorption of hydrogen in palladium using laser distillation},}\
  }\href@noop {} {\bibfield  {journal} {\bibinfo  {journal} {J. Chem. Phys.}\
  }\textbf {\bibinfo {volume} {131}},\ \bibinfo {pages} {084716} (\bibinfo
  {year} {2009})}\BibitemShut {NoStop}%
\bibitem [{\citenamefont {Tremblay}, \citenamefont {Monturet},\ and\
  \citenamefont {Saalfrank}(2010)}]{10:TMS:diss}%
  \BibitemOpen
  \bibfield  {author} {\bibinfo {author} {\bibfnamefont {J.~C.}\ \bibnamefont
  {Tremblay}}, \bibinfo {author} {\bibfnamefont {S.}~\bibnamefont {Monturet}},
  \ and\ \bibinfo {author} {\bibfnamefont {P.}~\bibnamefont {Saalfrank}},\
  }\bibfield  {title} {\enquote {\bibinfo {title} {Electronic damping of
  adsorbate vibrations at metallic surfaces},}\ }\href@noop {} {\bibfield
  {journal} {\bibinfo  {journal} {Phys. Rev. B}\ }\textbf {\bibinfo {volume}
  {81}},\ \bibinfo {pages} {125408} (\bibinfo {year} {2010})}\BibitemShut
  {NoStop}%
\bibitem [{\citenamefont {Tremblay}, \citenamefont {Monturet},\ and\
  \citenamefont {Saalfrank}(2011)}]{11:TMS:noau}%
  \BibitemOpen
  \bibfield  {author} {\bibinfo {author} {\bibfnamefont {J.~C.}\ \bibnamefont
  {Tremblay}}, \bibinfo {author} {\bibfnamefont {S.}~\bibnamefont {Monturet}},
  \ and\ \bibinfo {author} {\bibfnamefont {P.}~\bibnamefont {Saalfrank}},\
  }\bibfield  {title} {\enquote {\bibinfo {title} {The effects of
  electron--hole pair coupling on the infrared laser-controlled vibrational
  excitation of {NO} on {A}u(111)},}\ }\href@noop {} {\bibfield  {journal}
  {\bibinfo  {journal} {J. Phys. Chem. A}\ }\textbf {\bibinfo {volume} {115}},\
  \bibinfo {pages} {10698--10707} (\bibinfo {year} {2011})}\BibitemShut
  {NoStop}%
\bibitem [{\citenamefont {Tremblay}(2011)}]{tremblay2011laser}%
  \BibitemOpen
  \bibfield  {author} {\bibinfo {author} {\bibfnamefont {J.~C.}\ \bibnamefont
  {Tremblay}},\ }\bibfield  {title} {\enquote {\bibinfo {title} {Laser control
  of molecular excitations in stochastic dissipative media},}\ }\href@noop {}
  {\bibfield  {journal} {\bibinfo  {journal} {J. Chem. Phys.}\ }\textbf
  {\bibinfo {volume} {134}},\ \bibinfo {pages} {174111} (\bibinfo {year}
  {2011})}\BibitemShut {NoStop}%
\bibitem [{\citenamefont {Tremblay}, \citenamefont {F{\"u}chsel},\ and\
  \citenamefont {Saalfrank}(2012)}]{tremblay2012excitation}%
  \BibitemOpen
  \bibfield  {author} {\bibinfo {author} {\bibfnamefont {J.~C.}\ \bibnamefont
  {Tremblay}}, \bibinfo {author} {\bibfnamefont {G.}~\bibnamefont
  {F{\"u}chsel}}, \ and\ \bibinfo {author} {\bibfnamefont {P.}~\bibnamefont
  {Saalfrank}},\ }\bibfield  {title} {\enquote {\bibinfo {title} {Excitation,
  relaxation, and quantum diffusion of {CO} on copper},}\ }\href@noop {}
  {\bibfield  {journal} {\bibinfo  {journal} {Phys. Rev. B}\ }\textbf {\bibinfo
  {volume} {86}},\ \bibinfo {pages} {045438} (\bibinfo {year}
  {2012})}\BibitemShut {NoStop}%
\bibitem [{\citenamefont {F{\"u}chsel}\ \emph {et~al.}(2012)\citenamefont
  {F{\"u}chsel}, \citenamefont {Tremblay}, \citenamefont {Klamroth},\ and\
  \citenamefont {Saalfrank}}]{fuchsel2012selective}%
  \BibitemOpen
  \bibfield  {author} {\bibinfo {author} {\bibfnamefont {G.}~\bibnamefont
  {F{\"u}chsel}}, \bibinfo {author} {\bibfnamefont {J.~C.}\ \bibnamefont
  {Tremblay}}, \bibinfo {author} {\bibfnamefont {T.}~\bibnamefont {Klamroth}},
  \ and\ \bibinfo {author} {\bibfnamefont {P.}~\bibnamefont {Saalfrank}},\
  }\bibfield  {title} {\enquote {\bibinfo {title} {Selective {E}xcitation of
  {M}olecule-{S}urface {V}ibrations in {H}$_2$ and {D}$_2$ {D}issociatively
  {A}dsorbed on {R}u (0001)},}\ }\href@noop {} {\bibfield  {journal} {\bibinfo
  {journal} {Isr. J. Chem.}\ }\textbf {\bibinfo {volume} {52}},\ \bibinfo
  {pages} {438--451} (\bibinfo {year} {2012})}\BibitemShut {NoStop}%
\bibitem [{\citenamefont {Tremblay}(2013)}]{tremblay2013unifying}%
  \BibitemOpen
  \bibfield  {author} {\bibinfo {author} {\bibfnamefont {J.~C.}\ \bibnamefont
  {Tremblay}},\ }\bibfield  {title} {\enquote {\bibinfo {title} {A unifying
  model for non-adiabatic coupling at metallic surfaces beyond the local
  harmonic approximation: From vibrational relaxation to scanning tunneling
  microscopy},}\ }\href@noop {} {\bibfield  {journal} {\bibinfo  {journal} {J.
  Chem. Phys.}\ }\textbf {\bibinfo {volume} {138}},\ \bibinfo {pages} {244106}
  (\bibinfo {year} {2013})}\BibitemShut {NoStop}%
\bibitem [{\citenamefont {Serwatka}\ and\ \citenamefont
  {Tremblay}(2019)}]{Tremblay:2019b}%
  \BibitemOpen
  \bibfield  {author} {\bibinfo {author} {\bibfnamefont {T.}~\bibnamefont
  {Serwatka}}\ and\ \bibinfo {author} {\bibfnamefont {J.~C.}\ \bibnamefont
  {Tremblay}},\ }\bibfield  {title} {\enquote {\bibinfo {title} {{Stochastic
  wave packet approach to nonadiabatic scattering of diatomic molecules from
  metals}},}\ }\href {\doibase 10.1063/1.5092698} {\bibfield  {journal}
  {\bibinfo  {journal} {J. Chem. Phys.}\ }\textbf {\bibinfo {volume} {150}},\
  \bibinfo {pages} {184105} (\bibinfo {year} {2019})}\BibitemShut {NoStop}%
\bibitem [{\citenamefont {Ford}\ and\ \citenamefont
  {O'Connell}(1999)}]{Ford:1999}%
  \BibitemOpen
  \bibfield  {author} {\bibinfo {author} {\bibfnamefont {G.~W.}\ \bibnamefont
  {Ford}}\ and\ \bibinfo {author} {\bibfnamefont {R.~F.}\ \bibnamefont
  {O'Connell}},\ }\bibfield  {title} {\enquote {\bibinfo {title} {{Comment on
  ``Dissipative Quantum Dynamics with a Lindblad Functional''}},}\ }\href
  {\doibase 10.1103/PhysRevLett.82.3376} {\bibfield  {journal} {\bibinfo
  {journal} {Phys. Rev. Lett.}\ }\textbf {\bibinfo {volume} {82}},\ \bibinfo
  {pages} {3376--3376} (\bibinfo {year} {1999})}\BibitemShut {NoStop}%
\bibitem [{\citenamefont {Breuer}\ and\ \citenamefont
  {Petruccione}(2007)}]{Breuer:2007}%
  \BibitemOpen
  \bibfield  {author} {\bibinfo {author} {\bibfnamefont {H.~P.}\ \bibnamefont
  {Breuer}}\ and\ \bibinfo {author} {\bibfnamefont {F.}~\bibnamefont
  {Petruccione}},\ }\href@noop {} {\emph {\bibinfo {title} {{The Theory of Open
  Quantum Systems }}}}\ (\bibinfo  {publisher} {Oxford University Press,
  Oxford, UK},\ \bibinfo {year} {2007})\BibitemShut {NoStop}%
\bibitem [{\citenamefont {Dalibard}, \citenamefont {Castin},\ and\
  \citenamefont {M{\o}lmer}(1992)}]{dalibard1992wave}%
  \BibitemOpen
  \bibfield  {author} {\bibinfo {author} {\bibfnamefont {J.}~\bibnamefont
  {Dalibard}}, \bibinfo {author} {\bibfnamefont {Y.}~\bibnamefont {Castin}}, \
  and\ \bibinfo {author} {\bibfnamefont {K.}~\bibnamefont {M{\o}lmer}},\
  }\bibfield  {title} {\enquote {\bibinfo {title} {Wave-function approach to
  dissipative processes in quantum optics},}\ }\href@noop {} {\bibfield
  {journal} {\bibinfo  {journal} {Phys. Rev. Lett.}\ }\textbf {\bibinfo
  {volume} {68}},\ \bibinfo {pages} {580} (\bibinfo {year} {1992})}\BibitemShut
  {NoStop}%
\bibitem [{\citenamefont {Dum}, \citenamefont {Zoller},\ and\ \citenamefont
  {Ritsch}(1992)}]{dum1992monte}%
  \BibitemOpen
  \bibfield  {author} {\bibinfo {author} {\bibfnamefont {R.}~\bibnamefont
  {Dum}}, \bibinfo {author} {\bibfnamefont {P.}~\bibnamefont {Zoller}}, \ and\
  \bibinfo {author} {\bibfnamefont {H.}~\bibnamefont {Ritsch}},\ }\bibfield
  {title} {\enquote {\bibinfo {title} {Monte carlo simulation of the atomic
  master equation for spontaneous emission},}\ }\href@noop {} {\bibfield
  {journal} {\bibinfo  {journal} {Phys. Rev. A}\ }\textbf {\bibinfo {volume}
  {45}},\ \bibinfo {pages} {4879} (\bibinfo {year} {1992})}\BibitemShut
  {NoStop}%
\bibitem [{\citenamefont {Carmichael}(1993)}]{carmichael1993}%
  \BibitemOpen
  \bibfield  {author} {\bibinfo {author} {\bibfnamefont {H.}~\bibnamefont
  {Carmichael}},\ }\href@noop {} {\emph {\bibinfo {title} {An Open Systems
  Approach to Quantum Optics}}},\ edited by\ \bibinfo {editor} {\bibfnamefont
  {H.}~\bibnamefont {Araki}}\ (\bibinfo  {publisher} {Springer-Verlag,
  Berlin},\ \bibinfo {year} {1993})\BibitemShut {NoStop}%
\bibitem [{\citenamefont {M{\o}lmer}, \citenamefont {Castin},\ and\
  \citenamefont {Dalibard}(1993)}]{molmer1993monte}%
  \BibitemOpen
  \bibfield  {author} {\bibinfo {author} {\bibfnamefont {K.}~\bibnamefont
  {M{\o}lmer}}, \bibinfo {author} {\bibfnamefont {Y.}~\bibnamefont {Castin}}, \
  and\ \bibinfo {author} {\bibfnamefont {J.}~\bibnamefont {Dalibard}},\
  }\bibfield  {title} {\enquote {\bibinfo {title} {Monte {C}arlo wave-function
  method in quantum optics},}\ }\href@noop {} {\bibfield  {journal} {\bibinfo
  {journal} {J. Opt. Soc. Am. B}\ }\textbf {\bibinfo {volume} {10}},\ \bibinfo
  {pages} {524--538} (\bibinfo {year} {1993})}\BibitemShut {NoStop}%
\bibitem [{\citenamefont {Worth}\ \emph {et~al.}()\citenamefont {Worth},
  \citenamefont {Beck}, \citenamefont {J{\"a}ckle},\ and\ \citenamefont
  {Meyer}}]{mctdh:package}%
  \BibitemOpen
  \bibfield  {author} {\bibinfo {author} {\bibfnamefont {G.~A.}\ \bibnamefont
  {Worth}}, \bibinfo {author} {\bibfnamefont {M.~H.}\ \bibnamefont {Beck}},
  \bibinfo {author} {\bibfnamefont {A.}~\bibnamefont {J{\"a}ckle}}, \ and\
  \bibinfo {author} {\bibfnamefont {H.-D.}\ \bibnamefont {Meyer}},\ }\href@noop
  {} {}\bibinfo {howpublished} {The {MCTDH} {P}ackage, {V}ersion 8.2, (2000).
  H.-D. Meyer, {V}ersion 8.3 (2002), {V}ersion 8.4 (2007). Current version:
  8.4.18 (2019). 
  http://mctdh.uni-hd.de}\BibitemShut {NoStop}%
\bibitem [{\citenamefont {Gland}, \citenamefont {Sexton},\ and\ \citenamefont
  {Fisher}(1980)}]{Gland:1980}%
  \BibitemOpen
  \bibfield  {author} {\bibinfo {author} {\bibfnamefont {J.~L.}\ \bibnamefont
  {Gland}}, \bibinfo {author} {\bibfnamefont {B.~A.}\ \bibnamefont {Sexton}}, \
  and\ \bibinfo {author} {\bibfnamefont {G.~B.}\ \bibnamefont {Fisher}},\
  }\bibfield  {title} {\enquote {\bibinfo {title} {{Oxygen interactions with
  the Pt(111) surface}},}\ }\href {\doibase
  https://doi.org/10.1016/0039-6028(80)90197-1} {\bibfield  {journal} {\bibinfo
   {journal} {Surface Science}\ }\textbf {\bibinfo {volume} {95}},\ \bibinfo
  {pages} {587--602} (\bibinfo {year} {1980})}\BibitemShut {NoStop}%
\bibitem [{\citenamefont {Wurth}\ \emph {et~al.}(1990)\citenamefont {Wurth},
  \citenamefont {St\"ohr}, \citenamefont {Feulner}, \citenamefont {Pan},
  \citenamefont {Bauchspiess}, \citenamefont {Baba}, \citenamefont {Hudel},
  \citenamefont {Rocker},\ and\ \citenamefont {Menzel}}]{Menzel:1990}%
  \BibitemOpen
  \bibfield  {author} {\bibinfo {author} {\bibfnamefont {W.}~\bibnamefont
  {Wurth}}, \bibinfo {author} {\bibfnamefont {J.}~\bibnamefont {St\"ohr}},
  \bibinfo {author} {\bibfnamefont {P.}~\bibnamefont {Feulner}}, \bibinfo
  {author} {\bibfnamefont {X.}~\bibnamefont {Pan}}, \bibinfo {author}
  {\bibfnamefont {K.~R.}\ \bibnamefont {Bauchspiess}}, \bibinfo {author}
  {\bibfnamefont {Y.}~\bibnamefont {Baba}}, \bibinfo {author} {\bibfnamefont
  {E.}~\bibnamefont {Hudel}}, \bibinfo {author} {\bibfnamefont
  {G.}~\bibnamefont {Rocker}}, \ and\ \bibinfo {author} {\bibfnamefont
  {D.}~\bibnamefont {Menzel}},\ }\bibfield  {title} {\enquote {\bibinfo {title}
  {Bonding, structure, and magnetism of physisorbed and chemisorbed {O}$_2$ on
  {P}t(111)},}\ }\href {\doibase 10.1103/PhysRevLett.65.2426} {\bibfield
  {journal} {\bibinfo  {journal} {Phys. Rev. Lett.}\ }\textbf {\bibinfo
  {volume} {65}},\ \bibinfo {pages} {2426--2429} (\bibinfo {year}
  {1990})}\BibitemShut {NoStop}%
\bibitem [{\citenamefont {Steininger}, \citenamefont {Lehwald},\ and\
  \citenamefont {Ibach}(1982)}]{Steiniger:1982}%
  \BibitemOpen
  \bibfield  {author} {\bibinfo {author} {\bibfnamefont {H.}~\bibnamefont
  {Steininger}}, \bibinfo {author} {\bibfnamefont {S.}~\bibnamefont {Lehwald}},
  \ and\ \bibinfo {author} {\bibfnamefont {H.}~\bibnamefont {Ibach}},\
  }\bibfield  {title} {\enquote {\bibinfo {title} {{Adsorption of oxygen on
  Pt(111)}},}\ }\href {\doibase https://doi.org/10.1016/0039-6028(82)90124-8}
  {\bibfield  {journal} {\bibinfo  {journal} {Surface Science}\ }\textbf
  {\bibinfo {volume} {123}},\ \bibinfo {pages} {1--17} (\bibinfo {year}
  {1982})}\BibitemShut {NoStop}%
\bibitem [{\citenamefont {{D\"ubal}}\ and\ \citenamefont
  {Quack}(1984)}]{Quack:1984}%
  \BibitemOpen
  \bibfield  {author} {\bibinfo {author} {\bibfnamefont {H.-R.}\ \bibnamefont
  {{D\"ubal}}}\ and\ \bibinfo {author} {\bibfnamefont {M.}~\bibnamefont
  {Quack}},\ }\bibfield  {title} {\enquote {\bibinfo {title} {Tridiagonal
  {Fermi} resonance structure in the {IR} spectrum of the excited {CH}
  chromophore in {CF$_3$H}},}\ }\href@noop {} {\bibfield  {journal} {\bibinfo
  {journal} {J. Chem. Phys.}\ }\textbf {\bibinfo {volume} {81}},\ \bibinfo
  {pages} {3779--3791} (\bibinfo {year} {1984})}\BibitemShut {NoStop}%
\bibitem [{\citenamefont {Marquardt}\ \emph {et~al.}(1986)\citenamefont
  {Marquardt}, \citenamefont {Quack}, \citenamefont {Stohner},\ and\
  \citenamefont {Sutcliffe}}]{Marquardt:1986}%
  \BibitemOpen
  \bibfield  {author} {\bibinfo {author} {\bibfnamefont {R.}~\bibnamefont
  {Marquardt}}, \bibinfo {author} {\bibfnamefont {M.}~\bibnamefont {Quack}},
  \bibinfo {author} {\bibfnamefont {J.}~\bibnamefont {Stohner}}, \ and\
  \bibinfo {author} {\bibfnamefont {E.}~\bibnamefont {Sutcliffe}},\ }\bibfield
  {title} {\enquote {\bibinfo {title} {{Quantum-mechanical Wavepacket Dynamics
  of the {CH} Group in the Symmetric Top $\rm X_3CH$ Compounds using Effective
  Hamiltonians from High-resolution Spectroscopy}},}\ }\href {\doibase
  10.1039/F29868201173} {\bibfield  {journal} {\bibinfo  {journal} {J. Chem.
  Soc., Faraday Trans.~2}\ }\textbf {\bibinfo {volume} {82}},\ \bibinfo {pages}
  {1173--1187} (\bibinfo {year} {1986})}\BibitemShut {NoStop}%
\end{thebibliography}

%

\appendix

\section*{Appendix: }
\label{mu_nu_kappa}
\setcounter{equation}{0}
\renewcommand{\theequation}{A\arabic{equation}}
The aim of this derivation is to demonstrate that a thermalized ensemble of multi-dimensional eigenstate does not necessarily imply thermalization of the individual modes at the same temperature.
Our starting point is an incoherent ensemble of eigenstates ($\lvert i \rangle$) described by the  density matrix
\begin{eqnarray}
\rho = \sum_i p_i \lvert i \rangle \langle i \rvert
\end{eqnarray}
with populations distributed according to the Boltzmann law
\begin{eqnarray}
p_i = \frac{e^{-E_i/k_{\rm B}T}}{Z}
\end{eqnarray}
Here, $E_i$ is the energy of state $i$ and $Z$ is the partition function.
In general, we can write
\begin{eqnarray}
H\,\lvert i \rangle = E_i\,\lvert i \rangle
\end{eqnarray}
As a minimal model, we will study a system with two degrees of freedom. 
The hamiltonian thus takes the form
\begin{eqnarray}
H=H_1+H_2+V
\end{eqnarray}
where $H_1$ and $H_2$ are the zeroth-order hamiltonian of modes 1 and 2, respectively, and $V$ is the potential coupling between the two modes.
The individual mode hamiltonians can be used to defined a zeroth-order basis
from the stationary Schr\"odinger equation
\begin{eqnarray}
H_1 \lvert \alpha \rangle = \epsilon_{\alpha}^{(1)} \lvert \alpha \rangle ~~;~~ H_2 \lvert \beta \rangle = \epsilon_{\beta}^{(2)} \lvert \beta \rangle
\end{eqnarray}
where $\alpha$ and $\beta$ are zeroth order states with energy $\epsilon_{\alpha}^{(1)}$ and $\epsilon_{\beta}^{(2)}$, respectively.
The coupled eigenstates $\lvert i \rangle$ of the full hamiltonian can be written as linear combinations of Hartree products 
\begin{eqnarray}
\lvert i \rangle = \sum_{\alpha \beta} c_{i, \alpha \beta} \, \lvert \alpha \rangle \otimes \lvert \beta \rangle
\end{eqnarray}
This structure can be exploited to evaluate the reduced populations of each mode by taking the partial trace of the density matrix $\rho$ over the remaining degree of freedom (DOF).
 
\subsection*{Case 1} 

The simplest case is that of uncoupled modes, which implies that the eigenstates of the total hamiltonian can be written as a single Hartree product
with an energy defined from the single mode energies
\begin{eqnarray}\label{estates}
\lvert i \rangle = \lvert \alpha \rangle \otimes \lvert \beta \rangle ~~;~~ E_i=\epsilon_{\alpha}^{(1)} + \epsilon_{\beta}^{(2)} 
\end{eqnarray}
In such a case the partition function takes a factorizable form
\begin{eqnarray}
Z=\sum_{\alpha\beta}e^{-(\epsilon_{\alpha}^{(1)} + \epsilon_{\beta}^{(2)})/k_{\rm B}T}=
\Bigg(\sum_\alpha e^{-\epsilon_{\alpha}^{(1)}/k_{\rm B}T}\Bigg)
\Bigg(\sum_\beta e^{-(\epsilon_{\beta}^{(2)}/k_{\rm B}T}\Bigg) = Z_1Z_2
\end{eqnarray}
The associated population computed from the Boltzmann distribution becomes equally factorizable
\begin{eqnarray}
p_i = \frac{e^{-(\epsilon_{\alpha}^{(1)} + \epsilon_{\beta}^{(2)})/k_{\rm B}T}}{Z} =
\Bigg(\frac{e^{-\epsilon_{\alpha}^{(1)}/k_{\rm B}T}}{Z_1}\Bigg) \,  \Bigg(\frac{e^{-\epsilon_{\beta}^{(2)}/k_{\rm B}T}}{Z_2}\Bigg)
\end{eqnarray}
This simplification should lead to independent thermalization in each mode.

The reduced density matrix of a given mode can be obtained by tracing out one DOF, e.g.,
\begin{eqnarray}
\rho^{(1)} = Tr_{(2)} \left(\rho\right)
\end{eqnarray}
In the factorizable case, we can write 
\begin{eqnarray}
\rho = \sum_{\alpha \beta} \rho_{\alpha \beta} \lvert \alpha \rangle \langle \alpha \rvert \otimes \lvert \beta \rangle \langle \beta \rvert
\end{eqnarray}
where the element of the density matrix are the populations defined above, $\rho_{\alpha\beta}=p_i$.
Taking the partial trace over the second DOF leads to
\begin{eqnarray}
\rho^{(1)} &=& Tr_{(2)}\left(\sum_{\alpha \beta} \rho_{\alpha \beta} \lvert \alpha \rangle \langle \alpha \rvert \otimes \lvert \beta \rangle \langle \beta \rvert \right)\nonumber \\
&=&\sum_{\gamma} \sum_{\alpha \beta}\rho_{\alpha \beta} \lvert \alpha \rangle \langle \alpha \rvert \otimes \langle \gamma \lvert \beta\rangle \langle \beta\lvert\gamma \rangle \nonumber \\
&=& \sum_{\alpha} \left(\sum_{\beta} \rho_{\alpha \beta}\right) \lvert \alpha \rangle \langle \alpha \rvert
\end{eqnarray}
Since the populations $\rho_{\alpha \beta}$ are factorizable, we can rewrite
\begin{eqnarray}
 \sum_{\beta} \rho_{\alpha \beta} &=& \sum_{\beta} \frac{e^{-\epsilon_{\alpha}^{(1)}/k_BT}}{Z_1} \,\,  \frac{e^{-\epsilon_{\beta}^{(1)}/k_{\rm B}T}}{Z_2}\nonumber \\
 &=&  \frac{e^{-\epsilon_{\alpha}^{(1)}/k_{\rm B}T}}{Z_1}  \left(\frac{\sum_{\beta}e^{-\epsilon_{\beta}^{(2)}/k_{\rm B}T}}{Z_2}\right) \nonumber \\
 &=&  \frac{e^{-\epsilon_{\alpha}^{(1)}/k_{\rm B}T}}{Z_1}   = \rho_{\alpha}^{(1)}
 \end{eqnarray} 
 This leads to the following reduced density matrix 
 \begin{eqnarray}
 \rho^{(1)} = \sum_{\alpha} p_{\alpha}^{(1)} \lvert \alpha \rangle \langle \alpha \rvert
 \end{eqnarray}
 for which the populations follow a Boltzmann distribution, $p_{\alpha}^{(1)}=\frac{e^{-\epsilon_{\alpha}^{(1)}/k_{\rm B}T}}{Z_1}$.
Since the two DOF are thermalized independently, we can conclude that, when the hamiltonian is separable, i.e. the coupling vanishes, $V \rightarrow 0$, the reduced probability are distributed thermally, at the same temperature.
 
 \subsection*{Case 2} 
 
 In the more general case where the coupling does not vanish, thermalization of the individual modes is not necessarily observed at equilibrium.
 Let us first define the coupled density matrix from the coupled eigenstates in Eq.\,\eqref{estates}, that is
\begin{eqnarray}
\rho = \sum_i p_i \sum_{\alpha \beta \alpha^{\prime} \beta^{\prime}} c_{i, \alpha \beta} c_{i, \alpha^{\prime} \beta^{\prime}}^{\star} \lvert \alpha\rangle \langle\alpha^{\prime} \rvert \otimes \lvert \beta \rangle \langle \beta^{\prime} \lvert
\end{eqnarray}
Taking the partial trace of the coupled density matrix yields
\begin{eqnarray}
\rho^{(1)} = Tr_{(2)}\left(\rho\right) &=& \sum_i p_i \sum_{\gamma}\sum_{\alpha \beta \alpha^{\prime} \beta^{\prime}} c_{i, \alpha \beta} c_{i, \alpha^{\prime} \beta^{\prime}}^{\star} \lvert \alpha\rangle \langle\alpha^{\prime} \rvert \otimes \langle \gamma\lvert \beta \rangle \langle \beta^{\prime} \lvert \gamma \rangle \nonumber \\
&=&\sum_i p_i \sum_{\alpha \alpha^{\prime}\beta}c_{i,\alpha\beta}c^{\star}_{i,\alpha^{\prime}\beta}\lvert \alpha \rangle \langle \alpha^{\prime} \lvert \nonumber \\
&=& \sum_{\alpha\alpha^{\prime}}\left(\sum_{i\beta} p_i c_{i, \alpha \beta}c^{\star}_{i,\alpha^{\prime}\beta}\right)\lvert \alpha \rangle \langle \alpha^{\prime} \lvert \nonumber \\
&=& \sum_{\alpha\alpha^{\prime}}\rho_{\alpha\alpha^{\prime}}^{(1)}\lvert \alpha \rangle \langle \alpha^{\prime} \lvert 
\end{eqnarray}
This represents a fully coherent reduced density matrix,
in which all states $\lvert\alpha\rangle$ are coherently interacting with all other states $\lvert\alpha'\rangle$ of a given mode {\it via} the other DOF.
The diagonal elements of the reduced density matrix can be recast as
\begin{eqnarray}
\rho_{\alpha \alpha} = \sum_{i\beta} p_i c_{i, \alpha \beta}c^{\star}_{i,\alpha \beta} = \sum_i p_i\sum_{\beta} \lvert c_{i, \alpha \beta}\rvert^2 = \sum_i p_i w_{i,\alpha}
\end{eqnarray}
with weights $w_{i, \alpha}=\sum_{\beta} \lvert c_{i, \alpha \beta}\rvert^2$.
These diagonal elements are populations of the different states $\lvert\alpha\rangle$ In general, the population do not have to obey a Boltzmann distribution because the weight $w_{i, \alpha}$ are different for different values of $\{i, \alpha\}$, and they depend on the coupling strength $V$ and the other mode indirectly via the expansion coefficients $c_{i, \alpha \beta}$.

\end{document}